\documentclass[fleqn,usenatbib,preprint]{mnras}

\usepackage{newtxtext,newtxmath}

\usepackage[T1]{fontenc}
\usepackage{ae,aecompl}

\usepackage{graphicx}	
\usepackage{amsmath}	
\usepackage{amssymb}	
\usepackage[usenames, dvipsnames]{color} 
\usepackage{subfig}
\usepackage{float}
\bibpunct{(}{)}{;}{a}{}{}

\title
[Amplification of magnetic fields by CR currents]
{Amplification of perpendicular and parallel magnetic fields 
by cosmic ray currents}

\author[Matthews et al.]{J.~H.~Matthews$^1$, A.~R.~Bell$^2$, K.~M.~Blundell$^1$ and 
A.~T.~Araudo$^{3,1}$
\\$^1$University of Oxford, Astrophysics, Keble Road, Oxford, OX1 3RH, UK
\\$^2$University of Oxford, Clarendon Laboratory, Parks Road, Oxford OX1 3PU, UK
\\$^3$Laboratoire Univers et Particules de Montpellier CNRS/Universite de Montpellier, Place E. Bataillon, 34095 Montpellier, France
}

\date{Accepted 2017 April 07. Received 2017 March 06.}

\pubyear{2015}

\begin{document}
\label{firstpage}
\pagerange{\pageref{firstpage}--\pageref{lastpage}}
\maketitle

\begin{abstract}
Cosmic ray (CR) currents through magnetised plasma
drive strong instabilities producing amplification of the magnetic field.
This amplification helps explain the CR energy spectrum as well as 
observations of supernova remnants and radio galaxy hot spots. 
Using magnetohydrodynamic (MHD) simulations,
we study the behaviour of the non-resonant hybrid (NRH) 
instability (also known as the Bell instability)
in the case of CR currents perpendicular and parallel to the initial 
magnetic field. We demonstrate that extending simulations of the perpendicular case to 3D reveals a different character to the turbulence from that observed in 2D. Despite these differences, in 3D the perpendicular NRH instability still grows exponentially far into the non-linear regime with a similar growth rate to both the 2D perpendicular and 3D parallel situations.
We introduce some simple analytical models to elucidate the physical behaviour, 
using them to demonstrate that the transition to the non-linear regime 
is governed by the growth of thermal pressure inside dense filaments at 
the edges of the expanding loops. We discuss our results in the context of 
supernova remnants and jets in radio galaxies. Our 
work shows that the NRH instability can amplify
magnetic fields to many times their initial value 
in parallel and perpendicular shocks.
\end{abstract}




\def\kh{\boldsymbol{\hat{k}}}
\def\bh{\boldsymbol{\hat{b}}}
\def\jh{\boldsymbol{\hat{j}}}
\def\jcr{j}
\def\jcb{\boldsymbol{j} \times \boldsymbol{B}}
\def\jcbo{\boldsymbol{j} \times \boldsymbol{B_0}}
\def\ucb{\boldsymbol{u} \times \boldsymbol{B}}
\def\mhd{\textsc{mh3d}}
\def\jperpb{\boldsymbol{j} \perp \boldsymbol{B_0}}
\def\jparab{\boldsymbol{j} \parallel \boldsymbol{B_0}}
\def\bb{\boldsymbol}
\def\alfven{Alfv\'en}

\section{Introduction}

Cosmic rays (CRs) are detected on Earth with energies up to $10^{20}$~eV,
but the origin of the highest energy CRs is still unknown.
The CR energy spectrum is characterised by an $E^{-2.7}$ power 
law up to a spectral break at $\approx 10^{15}$~eV, 
commonly known as the knee, after which the spectrum steepens. 
There is now good evidence that CRs are 
accelerated up to, and 
possibly beyond, the knee in Supernova remnants 
\citep[SNRs;][]{allen1997,buckley1998,tanimori1998,vink2003}. 
The best candidate mechanism for accelerating 
high energy CRs is diffusive shock acceleration 
\citep[DSA;][]{axford1977,krymskii1977,bell1978a,blandford1978},
in which CRs are scattered multiple times across a 
shock front, successively gaining energy.
DSA produces an $E^{-2}$ power law spectrum, which is
consistent with that observed when escape losses in the 
Galactic disc are taken into account.

The characteristic maximum energy attainable by a CR undergoing DSA is set by the \cite{hillas1984} 
energy, which can be written as
\begin{equation}
T_H = 
Z~
\left( \frac{B}{\mathrm{\mu G}} \right)~
\left(\frac{u_s}{c}\right)~
\left(\frac{R}{\mathrm{kpc}}\right)~
0.9 \times 10^{18}~\mathrm{eV}
\label{eq:hillas}
\end{equation}
where $Z$ is the charge on the particle, $R$ is the size of the acceleration region, $u_s$ is the shock 
velocity and $B$ is the magnetic field. 
Since $u_s$ and $R$ are set from the 
parameters of the system in question, it becomes clear that the value of $B$ is 
crucial in estimating the maximum energy available to CR in a given astrophysical 
system. 

It is possible to estimate the magnetic field strength in a non-thermal
plasma from the observed synchrotron emission. In SNRs, X-ray synchrotron emission 
is confined to a thin rim close to the shock \citep[e.g.][]{hughes2000,long2003}. 
Associating the synchrotron cooling length with the thickness of these rims
allows the strength of the magnetic field to be inferred, 
resulting in estimates of $B$ roughly 100 times in excess of the 
expected ambient value of a few $\mu G$ 
\citep{vink2003,berezhko2003,berezhko2004,volk2005,ballet2006,cassam-chenai2007,uchiyama2007}. 
Similar estimates are obtained from observations of hot spots in radio 
galaxies \citep[][Araudo et al. in preparation]{araudo2015}, 
thought to be a potential source of ultra-high energy CRs 
\citep[UHECRs; e.g.][]{axford1994}. 
A mechanism to strongly amplify the magnetic field 
is therefore required to explain both the acceleration of the
highest energy cosmic rays and the observed non-thermal emission
from shocks.

CR themselves can cause fluctuations in 
the magnetic field via the resonant 
\citep{lerche1967,kulsrud1969,wentzel1974,skilling1975,skilling1975b, skilling1975c}
and non-resonant hybrid \citep[NRH; ][]{LB2000,bell2004,bell2005} 
instabilities; \cite{bell2004} showed that the latter causes strong
field amplification. The NRH instability is also 
known in the literature as the Bell instability,
and has been extensively studied using both magnetohydrodynamic 
\citep[MHD; ][]{bell2004,bell2005,zirakashvili2008,reville2008,beresnyak2014}
and particle-in-cell \citep[PIC; ][]{niemiec2008,RS2009,stroman2009}
simulations, as well as hybrid MHD-PIC treatments that
combine the two techniques \citep{LB2000,RB2012,RB2013,BSRG2013}.
The NRH instability is driven by a cosmic ray 
current, $\bb{j}$, which is introduced through an additional 
term in the MHD equations. This current drives exponential growth 
of the magnetic field, even when the perturbed field has grown stronger
than the initial uniform field, $\bb{B}_0$ 
(that is, once the instability has become non-linear).
Other instabilities are important for particle acceleration
and magnetic field amplification; for example, the 
\cite{weibel1959} instability can provide the turbulence needed
to scatter low energy CRs \citep{spitkovsky2008,sironi2011}, 
whilst other long-wavelength instabilities
\citep{drury1986,bykov2011,schure2011,reville2012} 
and vorticity created by density inhomogeneities 
\citep{giacalone2007,mizuno2014} 
may help perturb or amplify the magnetic field.

Much of the work on the NRH instability has
focused on the case in which $\bb{j}$ is parallel to the 
initial magnetic field, the situation in parallel shocks such 
as those thought to accelerate high energy particles in SNRs. 
However, perpendicular shocks are also common,
especially in relativistic cases when  the shock is always 
quasi-perpendicular. Highly relativistic shocks are unlikely 
to be the source of UHECR 
\citep[][Araudo et al. in preparation; Bell et al. in preparation]{lemoine2010,sironi2011,reville2014}, but 
kpc-scale, mildly relativistic shocks ($u_s \sim c/3$)
may permit acceleration up to $10^{20}$~eV if the value of
$B$ is high enough ($\sim 100 \mu \mathrm{G}$ from equation~\ref{eq:hillas}, 
which assumes Bohm diffusion). Bohm diffusion must apply so 
that high energy CRs do not escape upstream 
before being accelerated and also has some observational 
support in SNRs \citep{stage2006}.
The question of whether a perpendicular current 
can effectively amplify the ambient magnetic field is 
therefore crucial in understanding whether, for example, 
jets from active galactic nuclei can accelerate the highest 
energy CRs.

\cite{bell2005} initially gave the general form of the 
NRH dispersion relation -- which governs the growth of 
the instability -- for all orientations of $\bb{j}$ 
and $\bb{B_0}$. \cite{milos2006} built on this by
developing an analytical model for turbulence driven by
CR currents perpendicular to the initial magnetic field
in gamma-ray burst (GRB) afterglows, finding that the instability
may generate a large-scale magnetic field consistent with GRB
synchrotron emission. \cite{RS10} considered the perpendicular case in detail,
also deriving the dispersion relation and related 
growth rates, before presenting two-dimensional (2D) PIC simulations 
using both constant CR current and CR injection at a boundary. 
They found that, in the constant current simulation, 
the instability grew exponentially in 
time but could become saturated due to charge separation in the plasma
(see section~\ref{sec:bmax}). PIC simulations are powerful
tools for modelling the interaction of CRs with magnetised 
plasma and are essential for studying 
the early stages of instability growth and particle acceleration
as well as the `injection problem' 
\citep[e.g.][]{eichler1979,burgess1987,malkov1995,giacalone2005,martins2009,spitkovsky2008,sironi2011,caprioli2015}. 
However, they are much more performance intensive than
MHD codes and cannot yet be easily run in 
three dimensions \citep[3D; although see e.g.][]{RS2009, huntington2015}. 
In addition, the nature of PIC simulations 
usually requires adopting artificially low
proton-to-electron mass ratios, which can affect results \citep{ruyer2016}.
Furthermore, particle localisation into fluid elements is not fully accounted for.
Simple MHD codes such as those originally used by \cite{bell2004} 
do not have these limitations and can thus be used, 
complementarily to PIC studies, to examine the effect of considering 
the problem in 3D.

\begin{figure}
	\includegraphics[width=0.45\textwidth]{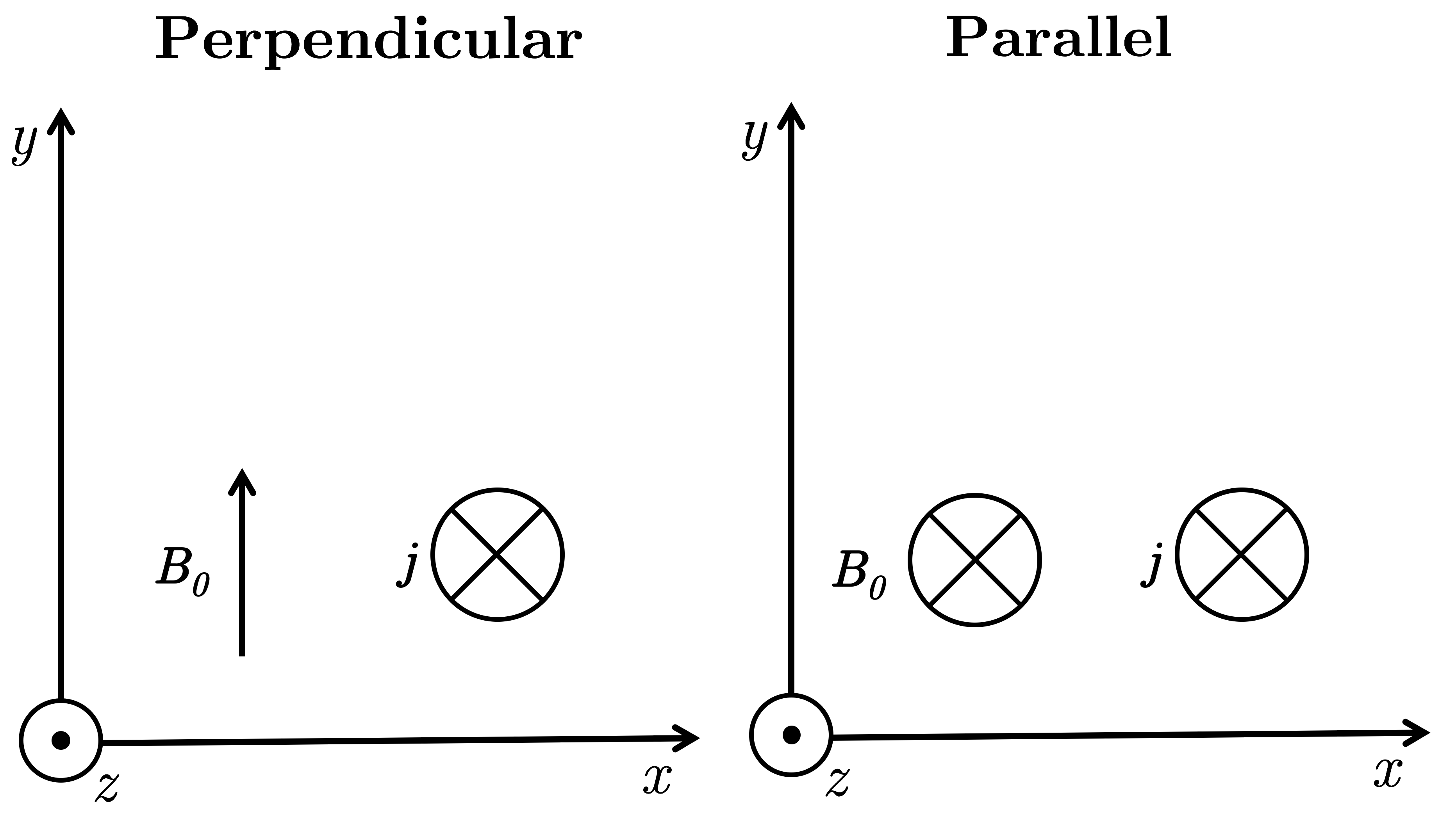}
    \caption
    {
    The initial configurations of the cosmic ray current and magnetic field
    for the perpendicular and parallel cases of the general, oblique
    NRH instability. The $\otimes$ and $\odot$ symbols indicate vectors into and out of
    the page, respectively, and the axes are oriented consistently with our 
    simulations, described 
    in section~\ref{sec:nonlinear_sims}. In the left hand panel, the CRs producing the current 
    drift perpendicular to both the initial magnetic field and the shock normal, whereas 
    in the right-hand panel the CRs drift along field lines in the upstream rest frame.
    In principle, this setup is general for imposed constant external current, $\bb{j}$.
    Note that RS10 refer to the perpendicular and parallel configurations 
    as the perpendicular current driven (PCDI) and cosmic-ray current driven (CRCD)
    instabilities, respectively.
    }
    \label{fig:geometry}
\end{figure}

In this study, we build upon the work of \cite{bell2004,bell2005} 
and \citet[][hereafter RS10]{RS10}, 
among others, by examining the linear and non-linear behaviour 
of the NRH instability in the case of CR currents 
perpendicular and parallel\footnote{Note that RS10 refer to 
the perpendicular and parallel configurations of the general, oblique 
NRH instability \citep{bell2005} as the cosmic-ray current driven (CRCD)
and perpendicular current driven (PCDI) instabilities.} 
to the initial magnetic field. The initial setup of the 
current and magnetic field for the two configurations is shown in
Fig~\ref{fig:geometry} and is discussed further in section~\ref{sec:nonlinear_sims}.
The paper is structured as follows. 
First, in section~\ref{sec:linear_theory}
we review the linear theory behind the NRH and give the
dispersion relations for parallel and perpendicular configurations.
In section~\ref{sec:nonlinear_sims},
we present our 2D and 3D MHD simulations and examine the growth 
of the instability in both cases. We also compare the results to 
earlier work on the NRH instability. In section~\ref{sec:discuss}, we
introduce some simple models for the turbulence in order to 
elucidate the plasma physics, which we use to 
explore the cause of the transition to the non-linear regime and
provide a summary of the main physical aspects of the NRH instability.
We discuss the overall implications of our work for CR acceleration
in section~\ref{sec:implications},
before concluding in section~\ref{sec:summary}.
We use SI units such that equations include $\mu_0$,
but adopt natural dimensionless units (i.e. $\mu_0=1$) 
for our simulations.

\section{Linear Theory}
\label{sec:linear_theory}

In ideal MHD, a fluid is described by the magnetic field, 
$\bb{B}$, the density, $\rho$, the pressure, $P$, and the local 
plasma velocity, $\bb{u}$. To define our problem, we also introduce 
a CR current, $\bb{j}$, which is carried by CRs with
number density $n_{\mathrm{cr}}$ and charge $e$. This CR current
induces an opposite return current of magnitude $j$ in the plasma; the turbulence
is then driven by a $-\bb{j}\times \bb{B}$ force.
For quasineutrality, the background plasma must also carry a 
charge density $-n_{\mathrm{cr}} e$. The velocity associated with the 
CR current depends on the geometry of the shock and the manner in which
the CRs drift in the vicinity of the shock front. 
The parallel shock case is described by \cite{bell2004,bell2005}, 
and focuses on the upstream restframe, in which the CRs have 
a drift velocity equal to the shock speed, $u_s$, giving 
$j=-n_{\mathrm{cr}} e u_s$. 
In this study, we also focus on 
both the downstream and upstream regions in a
perpendicular shock;
the application of the simulations to the actual 
shock regions is described in 
section~\ref{sec:implications}. In both the upstream 
and downstream regions, CRs drift
perpendicular to both the shock 
normal and the magnetic field with a velocity $v_d$, 
such that $j=-n_{\mathrm{cr}} e v_d$. 
CR currents are expected to exist both sides of a 
perpendicular shock boundary \citep{bell2011}.
A schematic showing the relative CR current and magnetic field configurations
for both parallel and perpendicular shocks is shown in Fig.~\ref{fig:geometry}.

In both parallel and perpendicular cases, 
we assume that the size of the MHD box is much
smaller than the CR Larmor radius, and we can thus treat the CRs
using a constant current density $\bb{j}$. This means
the basic MHD formalism is the same regardless of geometry and 
allows a completely general dispersion relation to be derived.
We thus write down the normal MHD equations in 
the upstream rest frame, but with an extra $-\jcb$ term
in the momentum equation to include the interaction of the 
CR current with the plasma: 
\begin{align}
\rho \frac{d \boldsymbol{u}}{dt} &= - \nabla P - \frac{1}{\mu_0}
\boldsymbol{B} \times (\nabla \times \boldsymbol{B}) - \jcb\
+ n_{\mathrm{cr}} e \ucb \ , \nonumber \\
\frac{\partial \boldsymbol{B}}{\partial t} &=
\nabla \times (\boldsymbol{u} \times \bb{B}) \ , \\
\frac{d \rho}{d t} &=
- \nabla \cdot ( \rho \boldsymbol{u})\nonumber \ .
\end{align}
These are the equations solved in our MHD simulations presented
in the next section. 
Although the electric force term, $n_{\mathrm{cr}} e \ucb$, 
can be important at low CR energies, it is
small when the CR driving the instability
have energies much greater than the thermal energies
\citep{bell2004}, so is neglected hereafter 
(see also section~\ref{sec:bmax}).
If we now separate each physical variable $\xi$
into a homogenous component, $\xi_0$, plus a perturbation, $\xi_1$,
we have 
$\bb{B}=\bb{B}_0 + \bb{B}_1$, 
$\bb{u}=\bb{u}_0 + \bb{u}_1$ and
$P=P_0 + P_1$. The zeroth-order velocity is time-dependent meaning
the plasma will, in general,
be subject to homogenous acceleration such that 
\begin{equation}
\frac{d \bb{u}_0}{dt} 
= - \frac{\bb{j} \times \bb{B}_0}{\rho_0}.
\end{equation}
This requires that the simulations are conducted in a non-inertial frame
when this acceleration is non-zero (see section~\ref{sec:nonlinear_sims}).
The first order MHD equations are
\begin{align}
\rho_0 \frac{d \bb{u}_1}{d t} &= 
- \nabla P_1
+ \frac{1}{\mu_0} (\nabla \times \bb{B}_1) \times \bb{B}_0 
+ \frac{\rho_1}{\rho_0} \bb{j} \times \bb{B}_0
- \bb{j} \times \bb{B}_1,\nonumber \\
\frac{\partial \boldsymbol{B}_1}{\partial t} &=
\nabla \times (\bb{u}_1 \times \bb{B}_0), 
\label{eq:first_order} \\
\frac{\partial \rho_1}{\partial t} &=
-\rho_0 (\nabla \cdot \bb{u}_1), \nonumber
\end{align}
where  we also define the sound speed
$c_s^2=\partial P / \partial \rho$. 
In the parallel case, the $\jcbo$ term can be eliminated from the above equations but this 
is not the case for $\jperpb$.

\cite{bell2004} showed that a strong instability is driven by the $\jcb$ terms in the above 
equations. The dispersion relation governing the growth of this instability can be derived by 
solving for $\bb{u}_1$ in the equations~\ref{eq:first_order}, before searching for
wave-like solutions with growth rate $\gamma$. This gives \citep{bell2005}

\begin{multline}
[\gamma^2 + (\kh\cdot\bh)^2 k^2 v_A^2]
[\gamma^4 + \gamma^2 k^2 (v_A^2 + c_s^2) + (\kh\cdot\bh) k^4 v_A^2 c_s^2]\\
~=\gamma_0^4~\Big \{ \gamma^2 + (\jh\cdot\kh) k^2 c_s^2\\
+ [(\kh\cdot\bh)^2 + (\kh\cdot\jh)^2 - 2(\kh\cdot\jh)(\bh\cdot\jh)(\kh\cdot\bh)]k^2 v_A^2 \Big  \} \,
\end{multline}
where $\gamma_0^4=(\bb{k} \cdot \bb{B}_0)^2 j^2 / \rho_0^2$,
$k$ is the wavenumber, $v_A = B_0/(\mu_0 \rho_0)^{1/2}$ is the \alfven\ speed, 
$c_s=(\partial P / \partial \rho)^{1/2}$ is the sound
speed and hats denote unit vectors. Up to this point the analysis is general
and independent of the orientation of $\bh$ and $\jh$. Rather than treat 
the general, oblique NRH instability
we now consider two cases: (i) the case when $\bh$ and $\jh$ are  
parallel, and (ii) the case when $\bh$ and $\jh$ are perpendicular. 
In both these cases, the fastest growth occurs in modes with wavevector 
$\bb{k}$ parallel to $\bb{B}_0$,
so we focus on these wavenumbers to derive the dispersion relations.

In the parallel case ($\jparab$), 
the dispersion relation then reduces to \citep{bell2005}
\begin{equation}
\left[\gamma^2 + k^2 c_s^2 \right]
\left[ (\gamma^2 + k^2 v_A^2)^2 - \gamma_0^4 \right] = 0 .
\end{equation}
Here, the two solutions are sound waves parallel to $B_0$, and 
\alfven-like transverse modes modulated by the 
magnetic tension. These transverse modes are purely growing if $k^2<\gamma_0^2/v_A^2$;
for larger $k$ the \alfven\ term starts to dominate and the modes
are oscillatory \citep{bell2004}. At small $k$ the wavelength becomes 
comparable to the Larmor radius of the CRs, meaning
they spiral along field lines and are inefficient in driving waves.
It can be shown that the maximum growth rate occurs at wavenumber 
$k_\mathrm{max} = 0.5~\mu_0 j / B_0$, which means the maximum growth rate 
is then 
\begin{equation}
\label{eq:gamma_max}
\gamma_{\mathrm{max}}=0.5~(\mu_0 / \rho_0)^{1/2}~j,
\end{equation}
as found by \cite{bell2004}. In the perpendicular case, the dispersion
relation differs and is given by
\begin{equation}
\left[\gamma^2 + k^2 v_A^2 \right]
\left[ (\gamma^2 + k^2 v_A^2)(\gamma^2 + k^2 c_s^2) - \gamma_0^4 \right] = 0,
\end{equation}
as also found by RS10. Here, the solutions are either \alfven\ waves, or 
compressional modes modulated by magnetic tension, magnetic 
pressure and
thermal pressure. The same maximum growth rate can be recovered 
by assuming that the sound speed is equal to the \alfven\ 
speed, which is reasonable provided
the sound speed is close to the typical turbulent velocity (RS10).
In this limit the maximum growth rate is again given by 
equation~\ref{eq:gamma_max}. The growth rate will actually decrease 
if $c_s$ increases beyond $v_A$; in other words, thermal pressure
can play a crucial role.



The dispersion relations and MHD equations show 
that there are two crucial differences between the 
$\jparab$ and $\jperpb$ limits. The first is the $\rho_1/\rho_0( 
\bb{j} \times \bb{B}_0)$ term in the momentum equation, which is non-zero when 
$\jperpb$; it follows that, in the perpendicular case, 
any physics that affects the density contrast in the plasma will also affect
growth of the instability. This includes forces operating along the 
axis parallel to $\bb{j}$. 
The second is the form of the dispersion relation; 
the presence of a $k^2 c_s^2$ term in the non-Alfv\'enic solution indicates 
that thermal pressure can also limit the growth rate, as can magnetic
tension in the field lines. Even in the cold plasma limit 
($P \rightarrow 0$), magnetic pressure gradients along the $z$-axis can affect the 
growth of the instability. Furthermore, the turbulent motion induced in 
the plasma from the $\jcbo$ term will lead to small-scale shocks that 
quickly increase the pressure far beyond its initial value. This is 
explored further in our MHD simulations in section~\ref{sec:nonlinear_sims}.


\section{MHD simulations in 2D and 3D}
\label{sec:nonlinear_sims}

To study the linear and non-linear behaviour of the instability, 
we conduct MHD simulations using the code \mhd\ 
\citep{LB1996, LB1997, LB2000, belllucek2001, bell2004}. 
\mhd\ is a 3-dimensional, Eulerian ideal MHD code 
that treats CRs as described in 
section~\ref{sec:linear_theory}, so that they are 
not collisionally connected to the plasma and only influence 
the dynamics through the return current that they induce, present in 
the MHD equations as $-\bb{j}$. The magnitude of 
$\bb{j}$ is constant and we adopt the convention of always 
orienting it along the negative $z$ axis in our simulations, 
as shown in Fig.~\ref{fig:geometry}. 
The energy equation is adiabatic,
with $P\propto \rho^{5/3}$, though we include viscous heating
such that kinetic energy can be converted to thermal energy 
in shocks. We have also adapted the code so it can be 
used in 2D, so as to reproduce the results
of RS10 and see if considering the problem in 
3D significantly affects the behaviour of the plasma. 
In 3D, the code is parallelised using OpenMPI \citep{openmpi}
in slices along the $z$ direction, whereas in 2D it is
parallelised in slices along the $y$ direction. 

\subsection{Simulation setup}

The plasma is initialised with a uniform magnetic field of 
strength $B_0$ oriented in either the $y$ or $z$ direction,
which correspond to the $\bb{j}\perp\bb{B}_0$ and 
$\bb{j}\parallel\bb{B}_0$ situations, respectively. The plasma
starts off at rest with a uniform density, $\rho_0$, and 
pressure, $P_0$, and is placed
in a non-inertial frame moving at velocity 
$\bb{u}_0 = t (\jcbo) / \rho_0$. The boundary conditions
are periodic, as in \cite{bell2004} and RS10. The magnetic field in the fluid 
is perturbed by taking the curl of a random number vector potential,
then normalising the amplitude of the fluctuations to $\delta B_0$.
The normalisation constraints are 
$\delta B_0 = 1/N \sum_N (\delta B_x^2 + 
\delta B_y^2 + \delta B_z^2)^{1/2}$ in 3D and 
$\delta B_0 = 3/(2N) \sum_N (\delta B_x^2 + \delta B_y^2)^{1/2}$ 
in 2D, where $N$ is the total number of grid cells. This 
ensures that the initial $B_x$ and $B_y$ components of the field 
have equal average magnitudes in 2D and 3D, but the initial magnetic 
energy density, $U_{\mathrm{mag}}$, 
and total magnitude, $B$, do not. However, our overall 
results are fairly insensitive to the starting
value of $B$ due to the rapid growth 
of the instability.

It is clear from the above MHD equations that, for a given random 
number seed and initial distribution of perturbation 
wavenumbers, $k$, the physical evolution of the instability is 
governed only by the relative orientation of $\bb{j}$ and $\bb{B_0}$, 
and by the relative values of the 
scalars $\rho_0$, $P_0$, $j$, $B_0$ and $\delta{B}_0$. 
The values of these quantities for our fiducial models are shown
in Table~\ref{tab:fiducial_params}; they are based on the  
simulations presented by \cite{bell2004}. 
We  specify whether the run in question was carried out in 2D 
or 3D and give the grid size in each case. 
We adopt a grid spacing of 
$\Delta x = \Delta y = \Delta z = 0.5$ 
and a minimum density, $\rho_{\mathrm{min}}=10^{-10}$, 
to prevent $c_s\rightarrow \infty$ and the time step, 
$\Delta t$, tending to zero. We have tested 
the sensitivity of our results to the adopted values of $\Delta x$, 
$\rho_{\mathrm{min}}$ and $N$ and verified that neither the
inferred growth rate nor the general character of the instability
is affected. Our simulations also reproduce behaviour found 
using both fixed-current PIC (RS10) and 
relativistic MHD \citep{beresnyak2014}
approaches, thereby independently verifying our results.

\begin{table}
	\centering
	\caption{Fiducial 
    model parameters used in the simulations.
    In each case, we set $P_0=1$, $j=2.512$
    and a cell size of $\Delta x =0.5$.}
	\label{tab:fiducial_params}
	\begin{tabular}{lcccccc} 
		\hline
		Run & 2D/3D? & Configuration & $B_0$ & $\delta B_0$ & Box Size\\
		\hline
		run2D & 2D & $\jperpb$ & 1 & 0.1 & $384^2$ \\
        run3Da & 3D  & $\jperpb$ & 1 & 0.1 & $384^3$ \\
        run3Db & 3D  & $\jparab$ & 1 & 0.1 & $384^3$ \\
		\hline
	\end{tabular}
\end{table}


\subsection{Simulation results}

\begin{figure*}
	\includegraphics[width=1.0\textwidth]{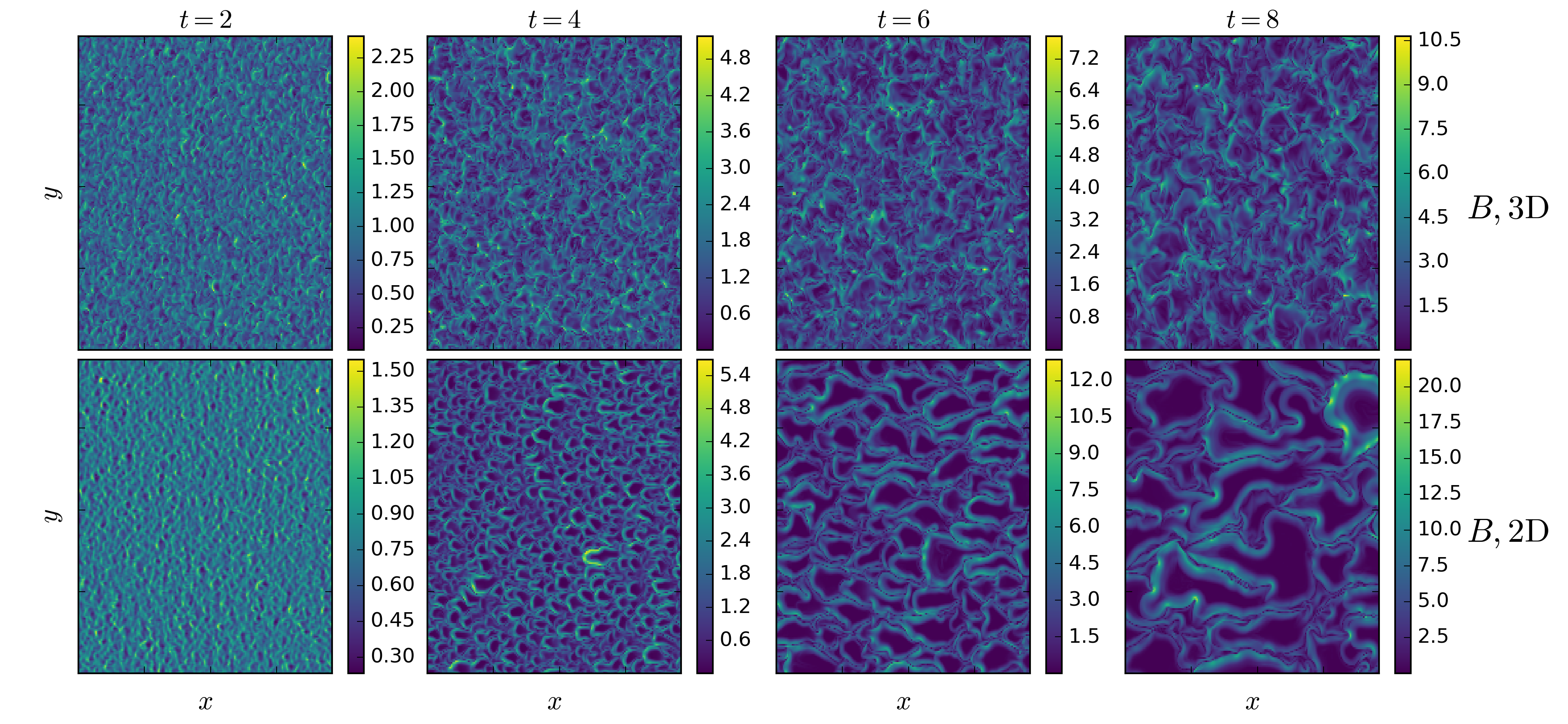}
    \caption
    {{\sl Magnetic field strength, perpendicular geometry:}
    Slices in the $xy$ plane of magnetic field strength, $B$,
    for the 3D (top; run3Da) and 2D (bottom; run2D) simulations. Time increases 
    to the right.
    $\bb{B}_0$ is oriented along the $y$ axis and $\bb{j}$
    is into the page. The bottom left-hand quadrant of the grid 
    is shown, with dimensions 192x192. 
    For the 3D simulation, the 2D slice is taken at 
    $z=192$.
    }
    \label{fig:xy_B}
\end{figure*}

\begin{figure*}
	\includegraphics[width=1.0\textwidth]{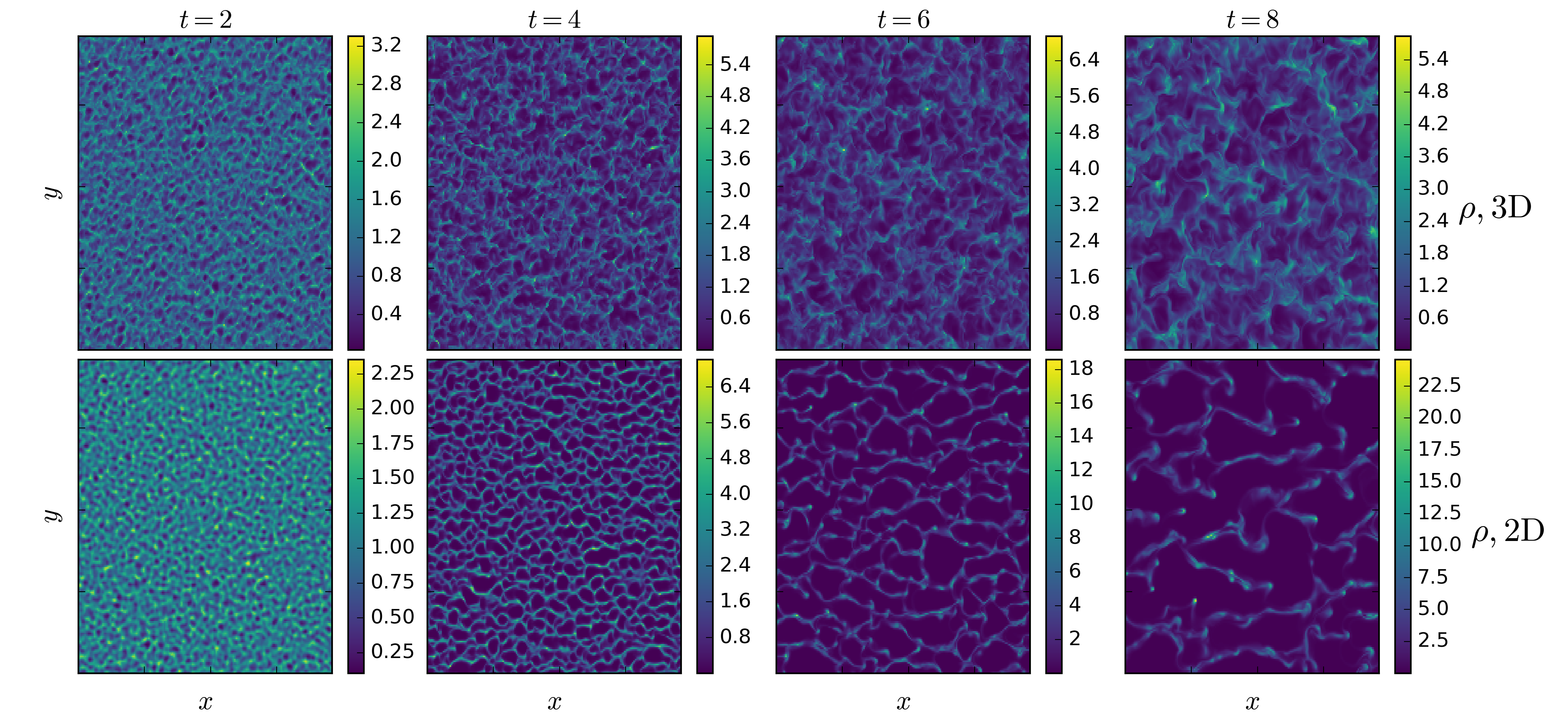}
    \caption
    {{\sl Density, perpendicular geometry:} 
    Slices in the $xy$ plane of density, $\rho$,
    for the 3D (top; run3Da) and 2D (bottom; run2D) simulations. Time increases 
    to the right. 
    $\bb{B}_0$ is oriented along the $y$ axis and $\bb{j}$
    is into the page. The bottom left-hand quadrant of the grid 
    is shown, with dimensions 192x192. 
    For the 3D simulation, the 2D slice is taken at 
    $z=192$.
    }
    \label{fig:xy_rho}
\end{figure*}

Fig.~\ref{fig:xy_B} shows the magnetic field strength in 
an $xy$ slice as a function of time,
from the 2D and 3D simulations. We find that the magnetic field 
grows exponentially from its initial seed value in both cases. 
Although there is a uniform density at first, the perturbed 
$(-\bb{j} \times \bb{B}_1)$ force immediately causes density fluctuations 
to emerge. This can be seen in the slices of density shown, 
again for 2D and 3D, in Fig.~\ref{fig:xy_rho}, and is also the reason for the small 
initial decrease in magnetic field strength. 
The $\rho_1/\rho_0 (\bb{j} \times \bb{B}_0)$ force then 
causes differential acceleration across the grid, with the 
result that overdense regions lag behind underdense regions. 
The enhances the density fluctuations and drives the instability.
Loops of high density and high magnetic field then expand 
exponentially, enveloping other loops and causing both the scale size
and energy associated with the turbulence to grow.

In 2D, our simulations mirror the behaviour found by RS10. 
We observe a similar anisotropy to the turbulence, which can be
understood in terms of the first-order momentum equation 
(equations~\ref{eq:first_order}; see also fig. 2 of RS10).
The $\rho_1/\rho_0 (\bb{j} \times \bb{B}_0)$ force acts to stretch 
out field lines along the $x$ axis, so larger values of $B_0$ tend
to enhance the anisotropy seen in Fig.~\ref{fig:xy_B}. 
In contrast, the $(-\bb{j} \times \bb{B}_1)$ force acts to stretch 
the loops of magnetic field in all directions. 
Thus, in general, the anisotropy to the turbulence increases 
with the ratio of $B_0$ to $\delta B_0$; 
the more ordered the field, 
the more there is a preferred direction to the turbulence.
As the perturbed field grows, the structure becomes more 
loop-like, with a series of low-density bubbles and 
high-density filaments. The magnetic field lines
are compressed near the edges of the expanding bubbles and the 
magnetic field is strongest in these locations. In 3D, the 
structure is similar to 2D at early times.
However, even in the linear regime
clear differences can be seen, namely that the density contrast
is smaller in the 3D case and the distinction between bubbles and filaments
is less pronounced. This is because pressure forces acting along the $z$-axis 
cause the low-density bubbles to be filled in, decreasing the density contrast 
in between the bubbles and filaments. Furthermore, in 3D modes with wavenumber
components in the $z$-direction can also contribute to the 
growth of the instability.

\begin{table}
	\centering
	\caption{Instability growth rates divided by the magnitude of the CR current density ($2.512$) for different magnetic field configurations,
    for both the linear and non-linear regimes. 
    The theoretical growth rate is roughly 
    equal to $\gamma_{\mathrm{max}}$ (equation~\ref{eq:gamma_max}).
    Note that $\gamma$ is defined as the growth rate for $B$, which means that 
    energy densities, as shown in Figs.~\ref{fig:energy_3d_by} and 
	\ref{fig:energy_2d} grow as $2\gamma$. The growth rates are obtained
    from fits to the $\Delta B|_{\mathrm{mean}}$ curve 
    (Figs.~\ref{fig:db_2dv3d} and \ref{fig:db_pp}) between $1<t<2.5$ for the
    linear regime and $8<t<10$ for the non-linear regime.
    }
	\label{tab:gammas}
	\begin{tabular}{lcccc} 
		\hline
		Field & Stage & $\gamma/j$~(theory) & $\gamma/j$~(2D) &$\gamma/j$~(3D)\\
		\hline
		$\jperpb$ & Linear & $\approx 0.5$ & 0.44 & 0.45 \\
        $\jperpb$ & Non-linear & - & 0.08 & 0.07 \\
        $\jparab$ & Linear & $\approx 0.5$ & - & 0.46 \\
        $\jparab$ & Non-linear & - & - & 0.07 \\
		\hline
	\end{tabular}
\end{table}

\begin{figure}
	\includegraphics[width=0.5\textwidth]{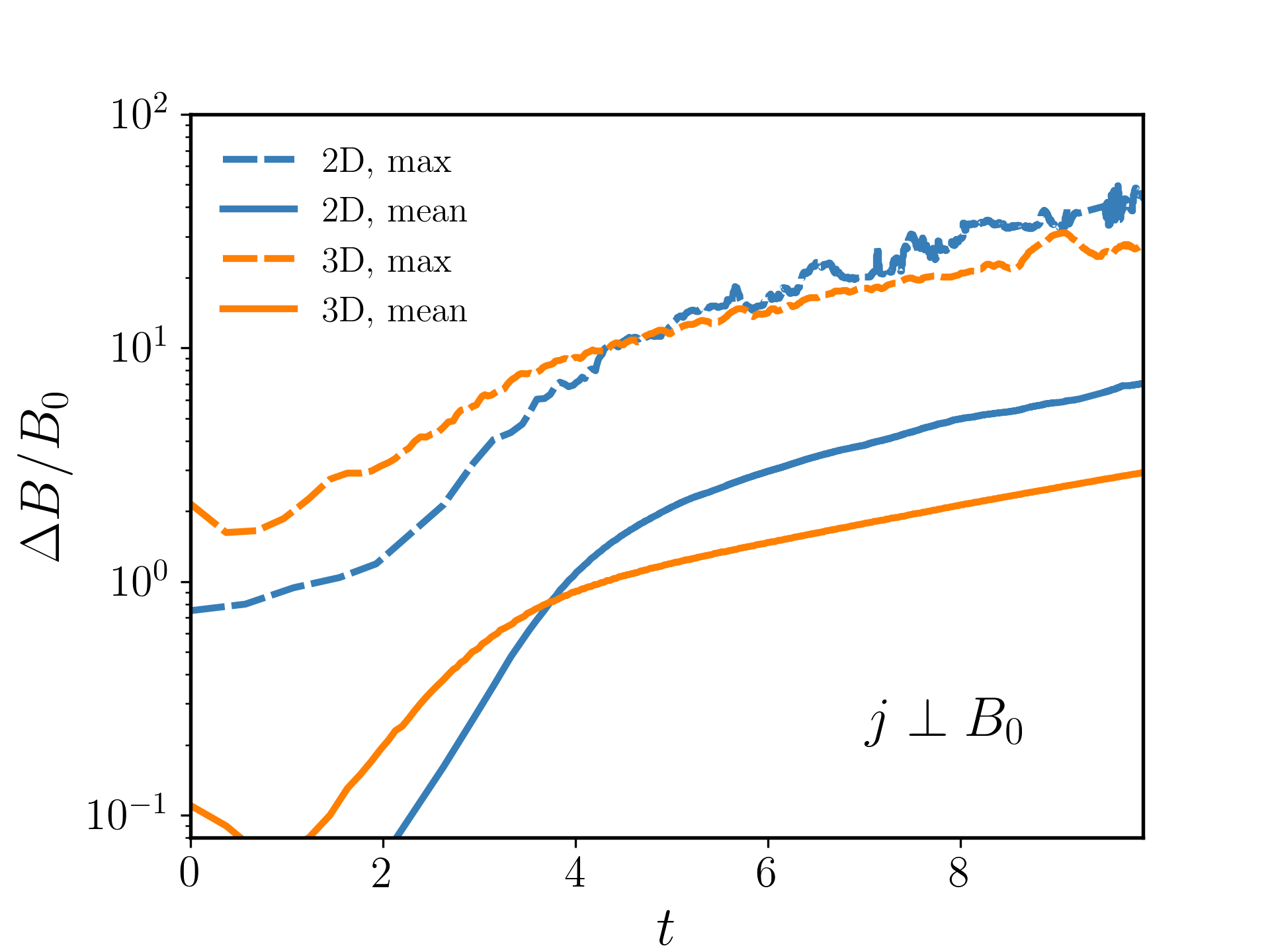}
    \caption
    {Magnetic field amplification factor, $\Delta B$,
    as a function of time in the 2D and 
    3D perpendicular simulations. The mean and maximum amplification factors 
    as defined in the text are shown with solid and dotted lines, respectively. 
    Values for $\gamma$ obtained from fits to these growth curves are shown 
    in Table~\ref{tab:gammas}.
    }
    \label{fig:db_2dv3d}
\end{figure}

\begin{figure}
	\includegraphics[width=0.5\textwidth]{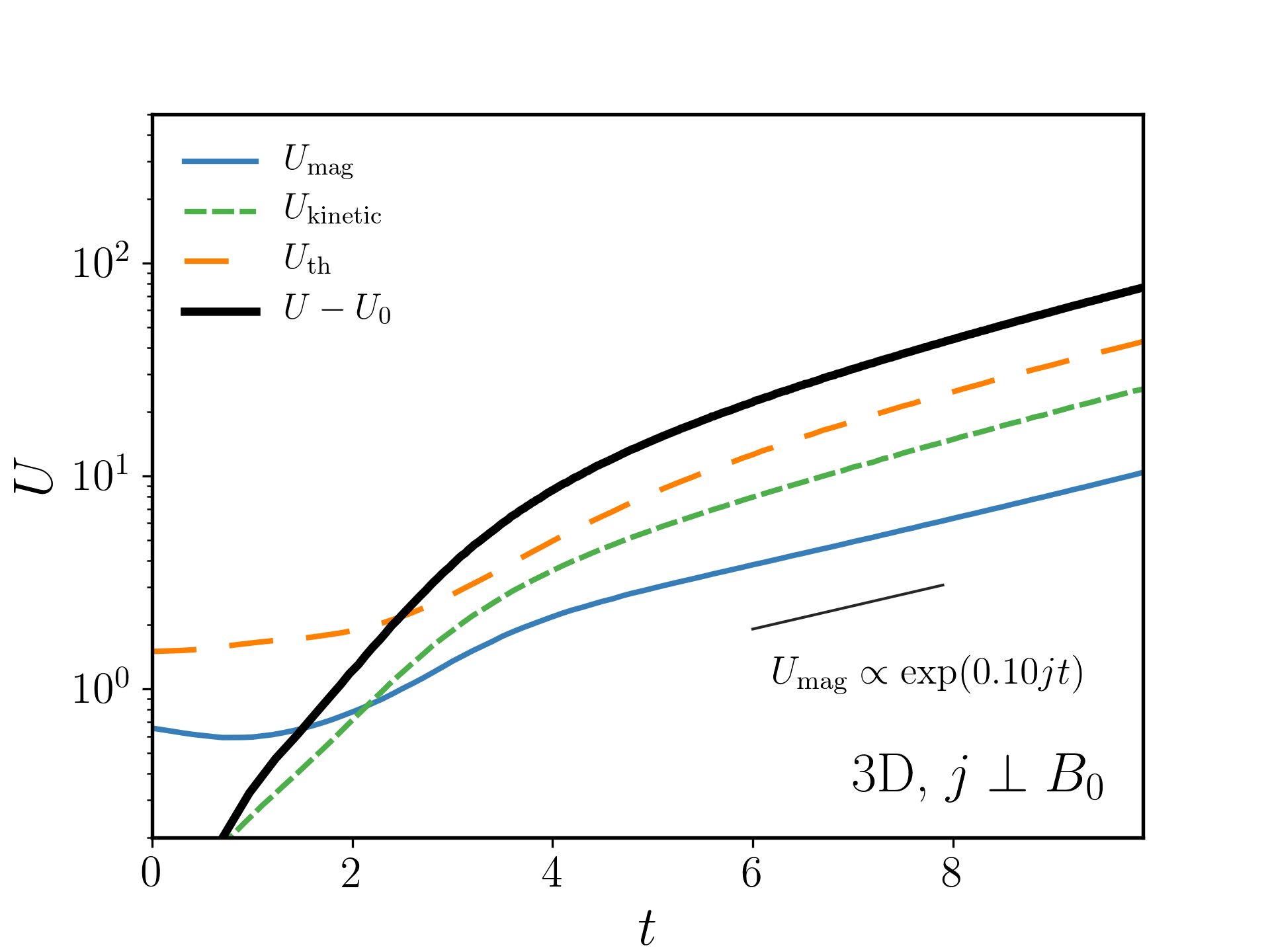}
    \caption
    {
    The evolution of the mean magnetic, kinetic and thermal energy densities
    in the 3D perpendicular simulation (run3Da), as well as the total energy 
    density increase. The energy grows at a rate of $\exp (2 \gamma t)$, 
    since $U\propto B^2$, hence the factor of 
    2 compared to Table~\ref{tab:gammas} and 
    Fig.~\ref{fig:db_2dv3d}.
    }
    \label{fig:energy_3d_by}
\end{figure}

\begin{figure}
	\includegraphics[width=0.5\textwidth]{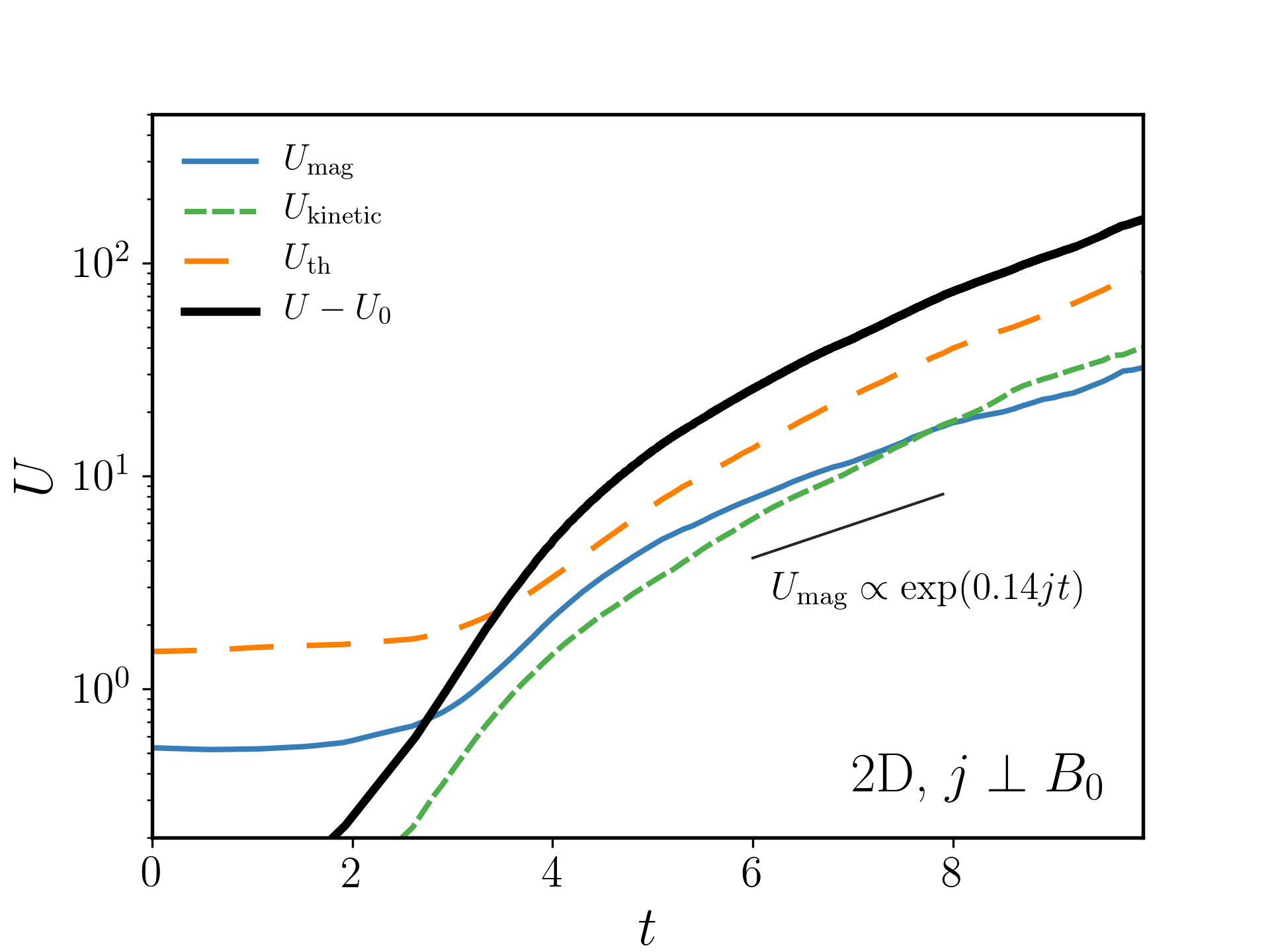}
    \caption
    {The evolution of the mean magnetic, kinetic and thermal energy densities in the 2D perpendicular simulation (run2D), as well as the total energy 
    density increase. The energy grows at a rate of $\exp (2 \gamma t)$ in energy units, hence the factor of 
    2 compared to Table~\ref{tab:gammas}.}
    \label{fig:energy_2d}
\end{figure}

\begin{figure}
	\includegraphics[width=0.5\textwidth]{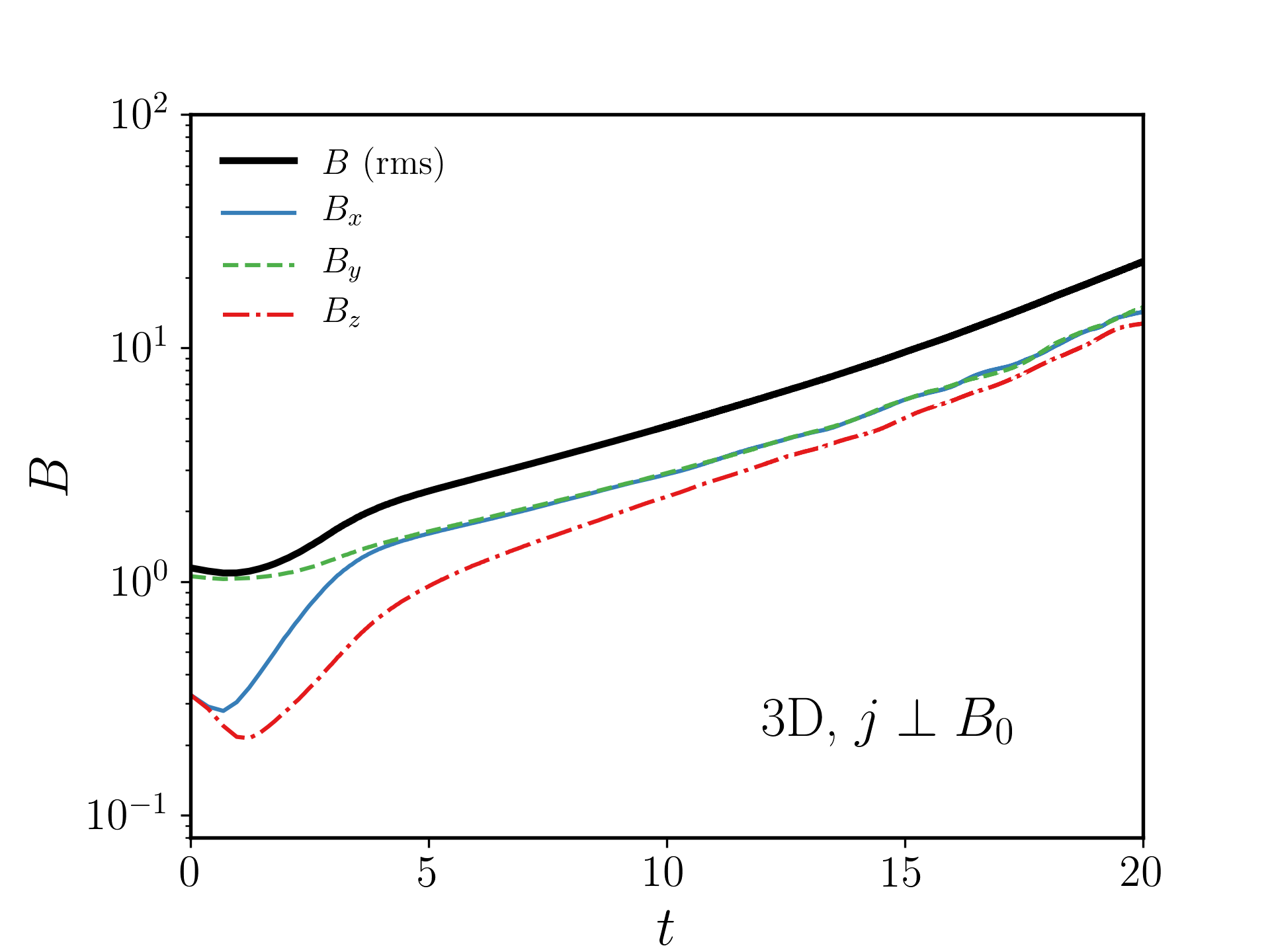}
    \caption
    {The magnetic field strength as a function of time for the 3D perpendicular simulation (run3Da). The rms values of the B-field in the $x,y$ and $z$ directions are also
    shown. The amplification continues exponentially to late times.}
    \label{fig:b_long}
\end{figure}

Fig.~\ref{fig:db_2dv3d} shows the evolution of the mean and 
maximum magnetic field amplification factors, which we define as
\begin{equation}
\Delta B|_{\mathrm{mean}} = 
\left[ \frac{1}{N} \sum_i^N B_i(t) \right] 
- B_0 \propto \exp(\gamma t)
\label{eq:deltab}
\end{equation}
and
\begin{equation}
\Delta B|_{\mathrm{max}} = B(t)|_{\mathrm{max}} - B_0,
\label{eq:deltab}
\end{equation}
respectively, where the sum is over all cells, $B_0=1$ is a constant 
and $B(t)|_{\mathrm{max}}$ is the maximum field strength in the simulation
box at time $t$. 
The magnetic field grows exponentially in the linear 
and non-linear regimes in both the 2D and 3D simulations, and the growth rate is reduced
by a similar factor once the instability becomes non-linear. 
The linear and non-linear growth rates, $\gamma_L$
and $\gamma_{NL}$, derived from an exponential fit in the two clearly 
delineated regimes, are given in Table~\ref{tab:gammas}. 
The thermal and kinetic energy density in the plasma also 
grow exponentially, as shown in Figs.~\ref{fig:energy_3d_by} and 
\ref{fig:energy_2d} where we show 
mean values of the various energy densities in the 
plasma over time, as well as the overall change in mean 
internal energy density, $U-U_0$. We also show the magnetic field
strength over time for the 3D simulation in Fig.~\ref{fig:b_long},
as well as the rms values of the components along each axis of the 
simulation box. Amplification continues to late times, with $B>10$,
and the $x$, $y$ and $z$ components converge to similar values.

\begin{figure*}
	\includegraphics[width=1.0\textwidth]{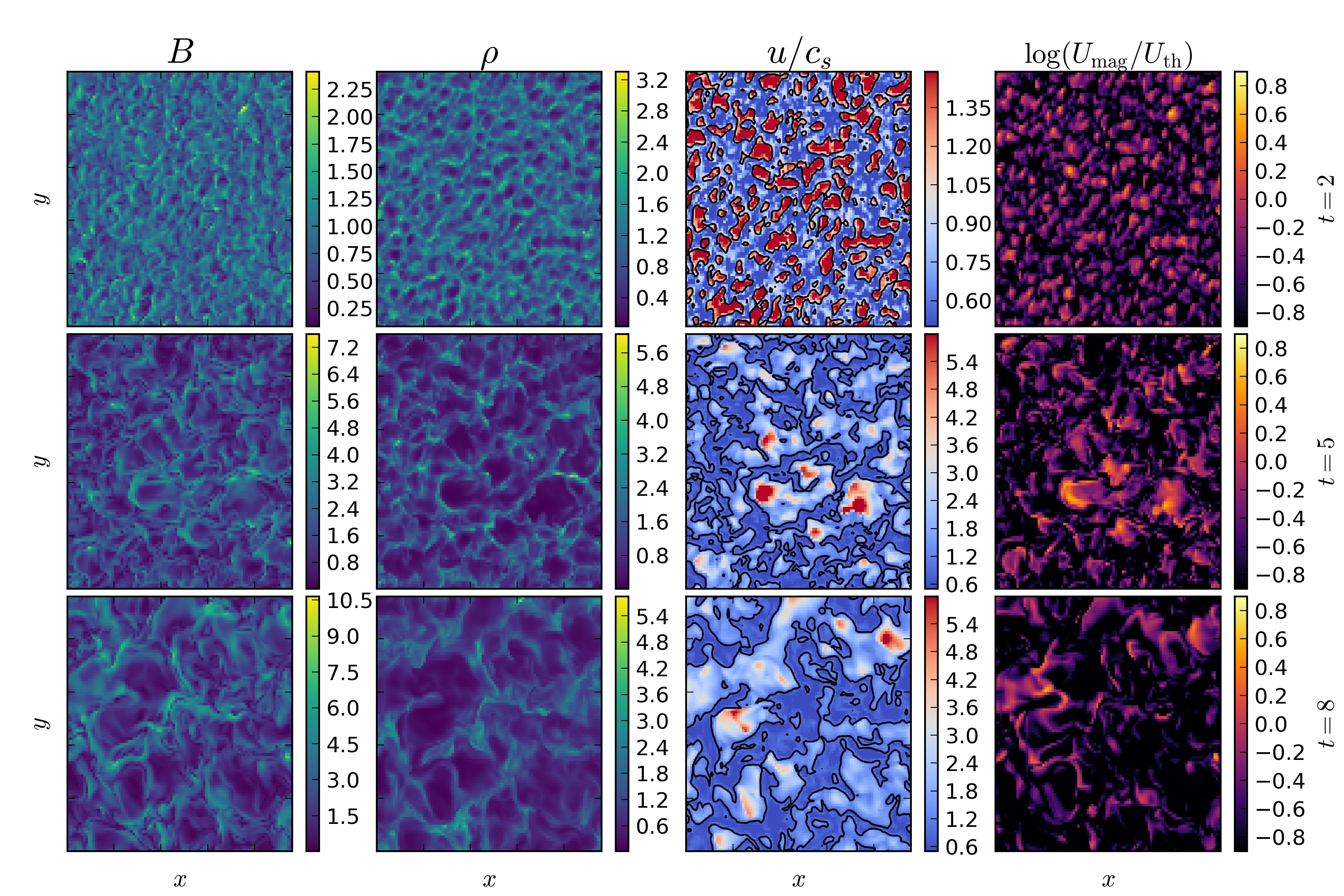}
    \caption
    {Slices in the xy plane from our fiducial 3D perpendicular simulation (run3Da) 
    at 3 time slices ($t=2,5,8$), showing four physical
    quantities of the plasma.
    These quantities are (left to right as labeled)
    magnetic field strength, density, the ratio of
    the plasma speed to the sound speed and the logarithm of the ratio of 
    the magnetic energy density to the thermal energy density. 
    In the third column, the black contour marks the Mach 1 surface.
    Loops of plasma expand in a qualitatively similar way to the 2D case, 
    forming regions of low density (`voids') surrounding by high density 
    filaments compressed by the magnetic field. 
    However, the presence of thermal and magnetic 
    pressure gradients allows some filling-in of the voids 
    from forces along the $z$-direction (parallel to $\bb{j}$). 
    The flow in the voids is largely supersonic,
    creating shocks that heat the plasma. The magnetic field strength is 
    highest close to the edge of the voids where magnetic 
    field lines press up against the high density filaments, 
    which limits the growth due to the thermal pressure.
    }
    \label{fig:four}
\end{figure*}

We show some of the physical aspects of the fiducial 3D perpendicular 
simulation in more detail in Fig.~\ref{fig:four}. 
We pick times that correspond roughly to before, during 
and after the linear to non-linear transition. The B-field and density are shown 
for reference, whilst the right-hand panels depict the ratio of plasma speed to
sound speed and the logarithm of the ratio of the
magnetic energy density to the thermal energy 
density. These plots reveal some key properties of the plasma. 
First, it is clear from Fig.~\ref{fig:four} 
that, broadly speaking, the flow is supersonic in the low density
voids, and subsonic in high density filaments. 
This leads to a series of small-scale shocks,
which heat the plasma and cause the rapid increase in thermal energy shown in 
Fig.~\ref{fig:energy_3d_by}. 
The thermal energy rises slightly later in the 2D 
case. The magnetic field lines are
compressed near the edges of the voids and the magnetic energy dominates
over thermal energy. However, in the regions of highest density, 
thermal energy is dominant; this is important in limiting the growth 
of the instability in the non-linear regime (see section~\ref{sec:transition}).

\subsection{Comparison to the parallel case}
\label{sec:parallel}

\begin{figure}
	\includegraphics[width=0.5\textwidth]{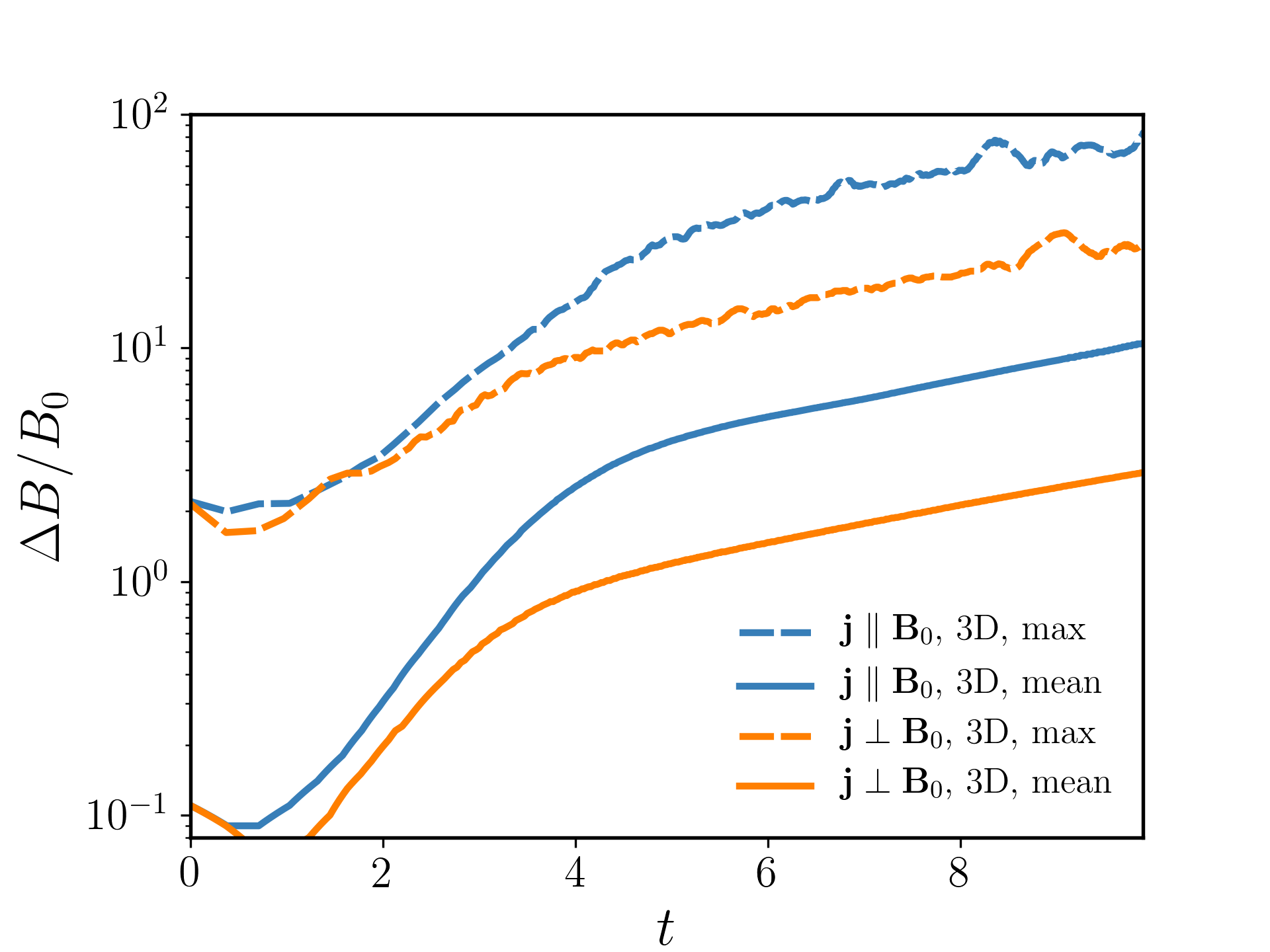}
    \caption
    {Magnetic field amplification factor as a function of
    time in the perpendicular (run3Da) and parallel (run3Db) current simulations. 
    The mean and maximum amplification factors 
    as defined in the text are shown with solid and dotted lines, respectively.
    Both simulations are in 3D.}
    \label{fig:db_pp}
\end{figure}

\begin{figure}
	\includegraphics[width=0.5\textwidth]{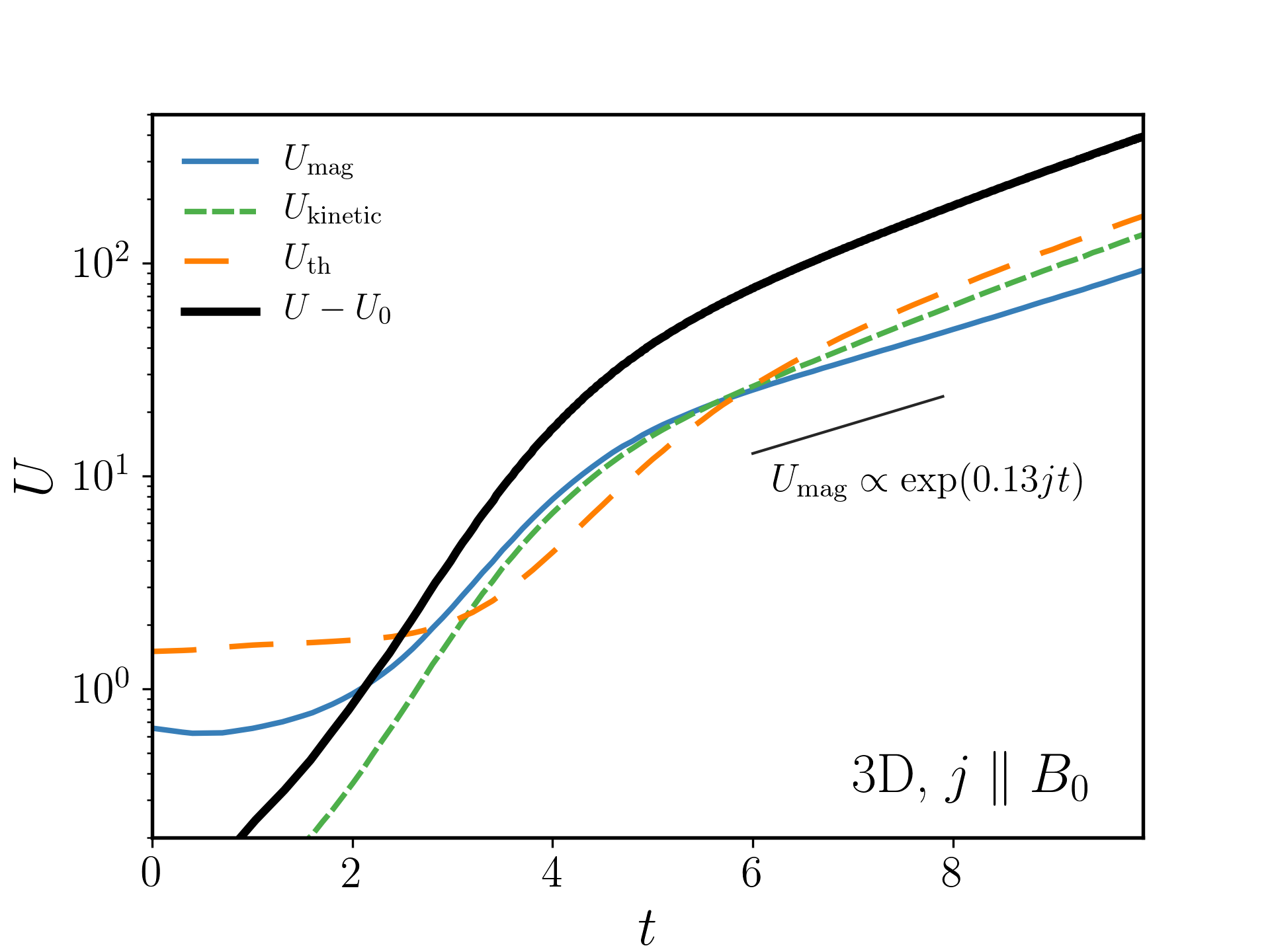}
    \caption
    {The evolution of the magnetic, kinetic and thermal energy in the plasma, 
    as well as the total energy density increase, for the parallel case (run3Db). 
    As figure~\ref{fig:energy_3d_by}, but for $\jparab$.}
    \label{fig:energy_para}
\end{figure}

The case in which a current parallel to $B_0$ drives the 
instability has been studied extensively in the literature 
\citep{bell2004,bell2005,zirakashvili2008,niemiec2008,RS2009,stroman2009,beresnyak2014}.
It is therefore instructive to compare the two physical pictures. 
The magnetic field amplification factor in the parallel and perpendicular cases 
is compared in Fig.~\ref{fig:db_pp}, while Fig.~\ref{fig:energy_para} shows
the evolution of the kinetic, thermal and magnetic energy densities in the 
parallel configuration. We recover a very similar growth rate 
in the non-linear regime to that reported by \cite{beresnyak2014} 
using an independent (relativistic) MHD code.

\begin{figure*}
	\includegraphics[width=1.0\textwidth]{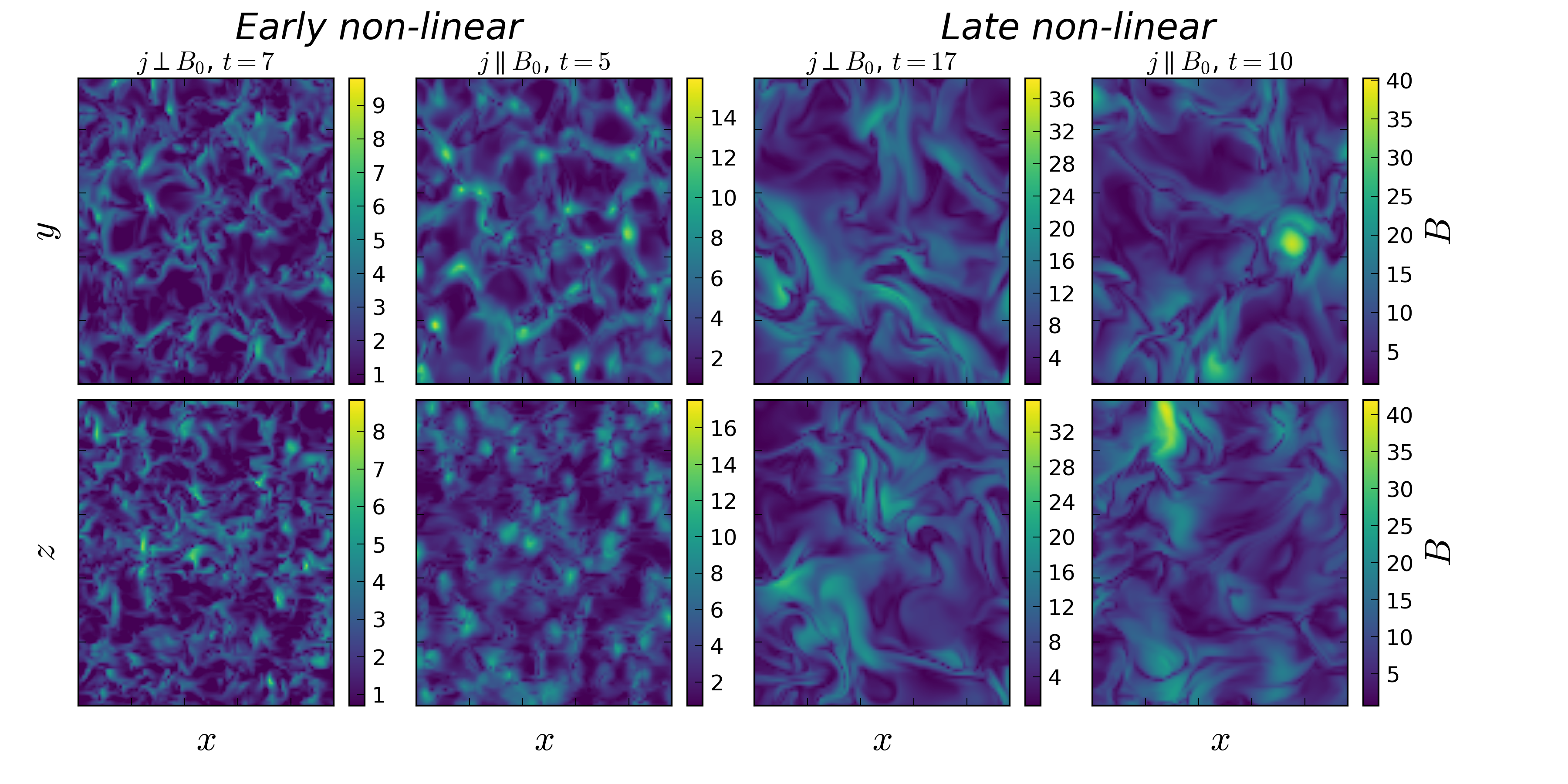}
    \caption
    {Slices in the $xy$ plane (Top) and $xz$ plane (Bottom) 
    of magnetic field strength $B$ for the marked geometries and times. 
    The four left-hand panels show the structure of the turbulence at early
    non-linear times; the plasma consists of tori with scale 
    $R(t) \sim \lambda e^{\gamma t}$
    in the $xy$ plane and width $\sim \lambda$ in the $z$ direction.
    At late times, shown in the four right-hand panels, the turbulence has mostly
    lost its preferred orientation and the typical scale-lengths 
    in each direction are similar.
    }
    \label{fig:zy_comp}
\end{figure*}

The growth rates (i.e. the {\em slopes} in Fig.~\ref{fig:db_pp}) 
are very similar in both the parallel and 
perpendicular cases. This is 
expected in the linear regime from the derived dispersion relations; 
at the maximally growing wavenumber, $k_{\mathrm{max}}$, 
the value of $\gamma$ is identical for $\jperpb$ and $\jparab$ 
if $v_A \approx c_s$, whilst $\gamma \rightarrow \gamma_0$ for 
small $k$. At late times, it is also unsurprising that
the growth rates are similar, as here $\Delta B / B_0 >1$ and 
the magnetic 
field does not retain its initial preferred direction. This can be seen in 
Fig.~\ref{fig:zy_comp}, which shows the magnetic field strength and density 
in both the $xy$ and $xz$ planes at both early and late times for the 
3D perpendicular and parallel simulations. 
At early times, the turbulence
exhibits a toroidal structure in which the typical scale-length
in the $xy$ plane is larger than the typical 
scale-length in the $z$-direction. 
The width of the filaments at the edge of the tori is noticeably larger
in the parallel geometry and there are  
a number of `hot spots' of strong magnetic field.
At late non-linear times, both the $xy$ anisotropy and 
toroidal structures have
largely disappeared; the turbulence has a similar 
structure in all directions 
regardless of the initial $B$-field configuration.
This behaviour can only be reproduced in 3D simulations, 
and can also be seen in Fig.~\ref{fig:b_long}, where 
the different rms fields in each direction
converge to the same value at late times.

Given the similarities in actual growth rates and late-time structure, 
the most important difference between the two configurations 
is thus the value of 
$\Delta B / B_0$ at which the growth rate is reduced from $\gamma_L$ 
to $\gamma_{NL}$. This transition happens at a lower magnetic field
strength in the perpendicular case. We discuss the reasons for this in
section~\ref{sec:transition}, then explore the astrophysical and observational 
consequences in section~\ref{sec:timescale}.


\section{Discussion}
\label{sec:discuss}

We have so far demonstrated a few key properties of the perpendicular
NRH instability in 3D and studied its character; importantly, we have
shown that while the density and B-field structure are different in 2D
and 3D, the instability continues to grow exponentially in both cases.
We also identified some key differences between both the perpendicular and parallel
instabilities and 2D and 3D simulations. In this section, we introduce some conceptual models
designed to elucidate the physics governing these differences -- as well as the similarities. 

\subsection{Rayleigh-Taylor-like description}
\label{sec:nonlinear_rt}

Although the perpendicular NRH instability is inherently 3D and 
non-linear, a simple, 2D analytical picture can nonetheless 
prove informative. Here, we apply a non-linear
analysis of the Rayleigh-Taylor (RT) instability \citep{ott1972}
to the perpendicular NRH instability. \cite{ott1972} 
found an exact, closed form solution for  RT growth in a thin layer.  

In RT, the problem involves 
a layer of fluid at $y=0$ which is supported against gravity 
($-g\hat{y}$) by a massless fluid at pressure $P_1$ in $y<0$. 
In $y>0$, there exists another massless fluid at pressure $P_2$.
Following \cite{ott1972}, 
we treat the problem in 2D and consider a surface element of the layer 
at $y=0,x=\xi_0$, with length $d\xi_0$ and mass $dm=w \sigma_0 d\xi_0$,
where $\sigma_0$ is the surface density.
and $w$ is the width in the $z$-direction.
We perturb the system at $t=0$. The force equation in this system is then 
\begin{equation}
\mathrm{d}m \frac{\partial^2 \vec{r}}{\partial t^2} = 
-g \mathrm{d}m \hat{y} - w (P_1 - P_2) 
\mathrm{d}\xi_0 \frac{\mathrm{d}\vec{r}}{\mathrm{d}\xi_0} \times \hat{z},
\end{equation}
where $\hat{x}$ and $\hat{z}$ are unit vectors.
In the NRH case, we replace the pressure gradient with a $j \times B$ 
force and associate the layer of fluid with a magnetic field line. 
We consider a non-inertial frame with the 
equilibrium condition that a psuedo-gravity of magnitude $g$ 
balances the homogenous acceleration, $jB_0/\rho_0$. We orient gravity along the $x$ direction
as in the simulations, such that $B_0$ is in the $y$-direction and 
$j$ is into the page along $z$. In this case, the equation of motion 
becomes
\begin{equation}
\mathrm{d}m \frac{\partial^2 \vec{r}}{\partial t^2} = -g dm \hat{x} - jB_0 
w \mathrm{d}h \mathrm{d}\xi_0 \frac{\mathrm{d}\vec{r}}{\mathrm{d}\xi_0} \times \hat{z},
\end{equation} 
where $\mathrm{d}h=\sigma_0/\rho_0$ is the thickness of the fluid layer
associated with the field line. Remembering that 
$\mathrm{d}m = w \sigma_0 \mathrm{d}\xi_0$ and 
imposing mass conservation, we can now separate into $x$ and $y$ 
and divide through by $\mathrm{d}m$ to obtain
\begin{equation}
\frac{\partial^2 x}{\partial t^2} = 
- g + (j B_0 / \rho_0) \frac{\mathrm{d}y}{\mathrm{d}\xi_0}
\end{equation}
\begin{equation}
\frac{\partial^2 y}{\partial t^2} = 
-g \frac{\mathrm{d}x}{\mathrm{d}\xi_0}
\end{equation}
where we know $g=jB_0 / \rho_0$. These equations 
allow the solution $y=\xi_0$, $x=0$, as required,
which is the equilibrium boundary condition.
There exists a special case of the solutions to these
equations in which the modes are purely growing such 
that, for a specific wavenumber $k$ we have
\begin{equation}
x(t) = A_0 \exp [ (k g)^{1/2} t ] \sin (k \xi_0),
\end{equation}
\begin{equation}
y(t) = \xi_0- A_0 \exp [ (k g)^{1/2} t ] \cos (k \xi_0).
\end{equation}
We can now write the growth rate as
\begin{equation}
\gamma = \sqrt{kg} = \sqrt{\frac{kjB_0}{\rho_0}},
\end{equation}
which is the characteristic growth rate 
found by \cite{bell2004} and obtained from the dispersion relations
in section~\ref{sec:linear_theory}.

\begin{figure}
	\includegraphics[width=0.5\textwidth]{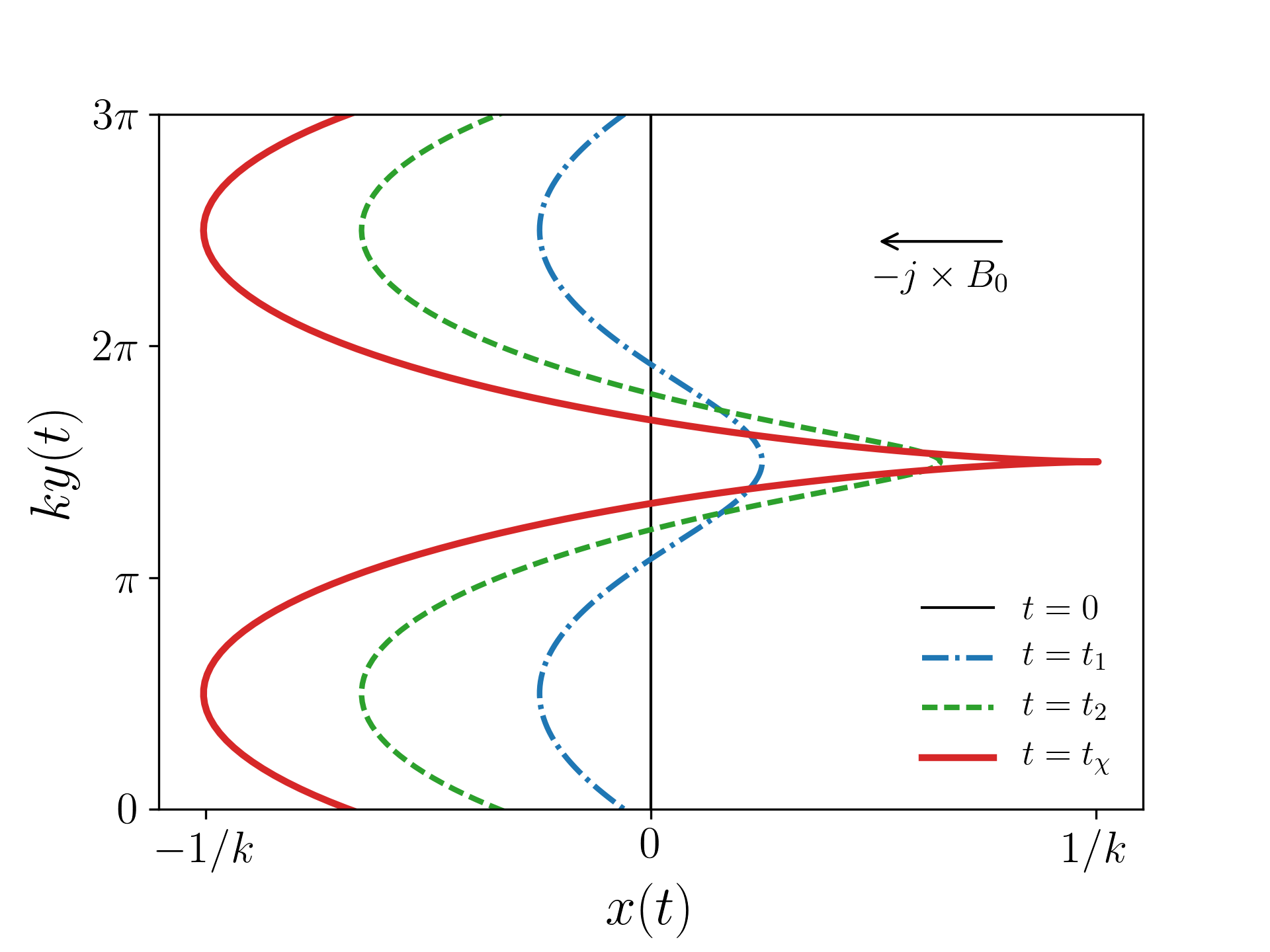}
    \caption
    {
    Schematic showing the cycloid form of a 
    field line deformed by a $-\bb{j}\times \bb{B}_0$ 	
    force at a few different times, as described in 
    section~\ref{sec:nonlinear_rt}. 
    The cycloids are constructed for arbitrary 
    parameter values.
    At early times, the field line forms a sinusoid 
    but is gradually distorted. 
    At some critical time $t_\chi$ the pressure in the
    `spike' of the cycloid limits the growth of the 
    instability and the analysis breaks down.
    In this plot the notional 
    $-\bb{j}\times \bb{B}_0$ force is oriented in 
    the same direction as the simulations so a 
    like-for-like 
    comparison can be made to Figs.~\ref{fig:geometry}, \ref{fig:xy_B} 
    and \ref{fig:xy_rho}.
    }
    \label{fig:cycloids}
\end{figure}

For a given mode of wavelength $\lambda = 2\pi / k$, the growth 
continues at this rate until the energy-containing scale 
is comparable to $\lambda$. 
The result of all this is that one finds that the functional
form of $x$ and $y$ is a cycloid evolving with time, as shown in 
Fig.~\ref{fig:cycloids} for arbitrary parameter values. 
Fig.~\ref{fig:cycloids}
is deliberately shown with the same orientation as the simulations
so a direct comparison can be made.
Early on, the field line 
is well described by a sinusoid. After a given time, 
the instability has grown so that the sinusoid is distorted, giving
the classical RT `bubbles and spikes' structure -- these are then 
analogous to the voids and filaments in our simulations. 
The instability grows exponentially because mass is displaced sideways
out of the bubbles, which in turn increases the $jB_0 / \rho$ acceleration.
Eventually, at $t\sim t_\chi$, the cycloid is so distorted that the pressure 
in the spikes limits the instability; at this point, the model
of differential acceleration due to density variations breaks down
and we must construct an alternative treatment. 

\subsection{Expanding loop model}
\label{sec:loop_mod}
\cite{milos2006} considered an analytical model for the perpendicular NRH instability
which the turbulence consists of a series of expanding loops. 
Similar models have been constructed by \citet[][for the parallel case]{bell2004,bell2005} 
and RS10. The growing modes are transverse in the parallel configuration, but compressional in
the perpendicular configuration. Loops in the $xy$ plane
are initially formed on the scale of the growing
modes, $\lambda$, but expand so that they extend to a radius $R$ at time $t$. 
Flux-freezing means that $B/(\rho R)$ is constant. Ignoring other
forces, the equation of motion is
\begin{equation}
\frac{d^2R}{dt^2} = jB/\rho = \frac{j B_0}{\rho_0 \lambda} R,
\end{equation}
with the result that $R(t) \sim \lambda \exp (\gamma t)$.
In reality, the expanding loops eventually have their growth limited
by a ram-pressure-like term, which will act on the cross-sectional area of 
the loop, $2 \pi R w$ where $w$ is the thickness in the $z$ direction.
This will happen at some critical value 
of $R$, which is geometry-dependent. 
The non-linear behaviour can then be thought
of as a competition between the growth of the loops due 
to the $-\jcb$ force 
and an external ram pressure acting on the loop. 
Under this assumption, the equation of motion becomes
\begin{equation}
\frac{d^2R}{dt^2} =  \frac{j B_0}{\rho_0 \lambda} R - 2 \pi R w\rho_{\mathrm{ext}} \left(\frac{dR}{dt}\right)^2,
\end{equation}
where, in general, $\rho_{\mathrm{ext}} \neq \rho_0$. The value of $B$ at 
which the transition to the non-linear regime occurs will therefore 
depend on the point at which this ram pressure becomes important, which
will occur once the loop has expanded to some critical radius, 
which we discuss in the next section.

\subsection{Transition to non-linear growth}
\label{sec:transition}
The transition to the non-linear regime occurs when 
the non-linear terms in equation~\ref{eq:first_order} 
become comparable to the linear ones. As the loops of 
magnetic field expand, they compress the plasma into dense 
filaments. After a time, the thermal pressure in these filaments becomes 
comparable to the $\jcb$ force causing their expansion 
and the thermal pressure starts to limit the growth of 
the instability. In general, this transition occurs at a time
$t_{\chi}$ such that the perturbed field, 
$B_1(t_\chi)=\chi B_0$.
We find $\chi \approx 1$ in 3D and 
$\chi \approx 2$ in 2D. This contrasts with the
value of $\chi \approx 3$ for the parallel
case found here and by \cite{bell2004} and \cite{beresnyak2014},
noting that the turnover will occur around $\chi^2$ in 
energy units.

The transition to the non-linear regime can be understood
in terms of the models described in sections~\ref{sec:nonlinear_rt}
and \ref{sec:loop_mod}. The key aspect of the instability that affects the 
value of $\chi$ is the typical size of the loops when the
ram pressure term becomes important. The geometry of the instability
is critical here; in the perpendicular case, the modes are compressional 
and $\kh$ (for $\gamma_{\mathrm{max}}$) 
is in the plane of the loops. As a result, the limiting scale 
is roughly $\lambda=2\pi/k$. 
This also follows from the RT analysis of 
section~\ref{sec:nonlinear_rt}. However, in the parallel case the 
loop expansion is transverse to the direction of $\kh$, 
and we can instead think of the critical scale as 
a `transverse coherence length' $R_C$, which has no knowledge of 
$\lambda$. The fastest growing modes have $\kh$
parallel to $\bb{B}_0$, so the spatial
variations perpendicular to $\bb{B}_0$ are small and 
$R_C > \lambda$. As such the different in values of 
$\chi$ can be attributed to the ratio $R_C/\lambda$.  

\subsection{A summary of the NRH instability in multiple geometries}

Since the original discovery papers by \cite{LB2000} and 
\cite{bell2004}, there have 
been a series of valuable contributions which have helped to expand our 
understanding of the physics and astrophysical significance of the 
NRH instability 
\citep[e.g.][]{bell2005,zirakashvili2008,niemiec2008,stroman2009,
RS2009,RS10,rogachevskii2012,beresnyak2014}.
Informed by our new results and this previous work, 
we briefly summarise what we consider 
the main physical characteristics of the instability.

The NRH instability can be broadly separated 
into three {\em geometries} (parallel, 2D perpendicular and 3D perpendicular)
and three time {\em regimes} (linear, early non-linear and late non-linear). 
For a given set of initial physical conditions, the geometry is 
determined by the alignment of $j$ and $B_0$, as well as the number of dimensions 
considered, while the regime is just determined by the time $t$.
The linear behaviour of the instabilities can be derived from the same, general, dispersion relation
and MHD equations, so the instabilities can be thought of as variants of the same,
generally oblique, instability.
However, there are clear differences in behaviour, and the actual forces driving and limiting
the instability are slightly different in each case. For clarity, we summarise the behaviour of
the instability within each geometry and regime here, with the aid of 
the RT and loop models discussed above which show that the growth will 
be exponential if $B / \rho$ increases in the loop.

\begin{itemize}
\item Linear parallel geometry: The instability does not enhance density fluctuations,
so $\rho$ is roughly constant. However, $B$ is the perturbed field and so increases in proportion
with the amplitude, leading to exponential growth.

\item Early non-linear parallel geometry: If we assume a cylindrical loop geometry in which $B / (\rho R)$
is constant, then we have $B\propto R$ and a continuing exponential growth.

\item Linear perpendicular geometry: Rayleigh-Taylor-like exponential growth -- differential 
sideways acceleration enhances density fluctuations, which in turn increases the acceleration
of the loops. Similar behaviour in 2D and 3D, but the transition to the non-linear regime occurs 
at different times.

\item Early non-linear perpendicular geometry, 2D: The mass displacement into `spikes' or `filaments'
of high density continues, creating large voids (bubbles) of low density. 
Bubbles still expand exponentially but are limited by the pressure from spikes.

\item Early non-linear perpendicular geometry, 3D: As in 2D, but pressure induces movement 
along the $z$ axis and allows infilling of the bubbles. This causes 
the $\rho$ in $B/\rho$ to 
increase and means that the point at which ram pressure and  
$jB/\rho$ become comparable occurs sooner, resulting in a lower value of 
$\chi$.

\item Late-time non-linear behaviour: At late times, all geometries converge to similar behaviours;
loop expansion limited by ram pressure. The amplified field is far greater than the initial field, 
and there is no preferred direction to the turbulence. Growth rates are similar, and we always 
have $U_{\mathrm{th}}>U_k>U_{\mathrm{mag}}$. 
\end{itemize}

\section{Overall implications for CR acceleration}
\label{sec:implications}

Although SNRs are widely thought to accelerate CRs up to
the knee in the CR energy spectrum, the origin of the 
highest energy CRs is still unknown. We therefore
explore the timescales and maximum field strength 
associated with the NRH instability (with both parallel
and perpendicular configurations), and examine whether they
satisfy the requirements for efficient CR acceleration to
high energies. The general principles behind this 
discussion will be described in more detail by
Bell et al. (in preparation), although there the discussion will be 
focussed on highly relativistic shocks.

\subsection{Maximum magnetic field strength}
\label{sec:bmax}
In our simulations, the magnetic field keeps growing as 
$B \propto \exp (\gamma_{NL} t)$, where $\gamma_{NL} \approx 0.1$.
There is no mechanism to saturate the instability in our ideal 
MHD approach. In reality, our treatment breaks down once the size of 
the turbulence becomes comparable to the Larmor radius of 
the CRs driving the instability. 
At that point the CRs start to move along field lines and kinetic 
effects become important, preventing further amplification.
The highest energy CRs are responsible for driving
the largest scale turbulence and thus set the maximum magnetic field
strength, $B_{\mathrm{sat}}$. These CRs, with Larmor radius $r_{g}$ 
and energy $T_{\mathrm{max}}$, drive the instability provided that
$B/r_{g} \sim \mu_0 j$, which is the condition for the CRs to exert 
a $\bb{j} \times \bb{B}$ force that overcomes the tension in the
magnetic field lines on a scale $r_{g}$.
The energy density of the CRs, $U_\mathrm{cr}$ is related to the
CR current by $U_\mathrm{cr} v_d \sim j T_{\mathrm{max}}$. 
Since $r_{g} = T_{\mathrm{max}} / (e c B)$, we then obtain
\begin{equation}
B_{\mathrm{sat}}^2 \sim \mu_0~(v_d / c)~U_{\mathrm{CR}},
\label{eq:bsat}
\end{equation}
for the saturated magnetic field strength \citep{bell2004}. 
This equation is very similar to that derived from equipartition 
arguments, but the above limit will apply at lower values of 
$B_{\mathrm{sat}}$ in general; this is because (i) generally $v_d < c$, 
and (ii) any steepening in the CR energy spectrum 
will mean the bulk of the energy is contained in CRs that cannot 
drive turbulence on the largest scales.

RS10 also showed that a number of other factors can 
influence the saturation point of the instability. One of 
these is charge separation in the plasma, something which is
not dealt with in our approach. 
Adopting $n_{\mathrm{cr}}/n_i = 0.01, 0.04$, RS10 show that, as the 
negatively charged loops expand, this focuses CR particles into 
the voids between the loops and induces
an electric field in the loops themselves, which opposes the 
$-\jcb$ force and limits their expansion. The focusing of CR charge into
the voids between loops was also found by \cite{bell2005} and \cite{RB2013}.
This electric field comes 
from the $n_{\mathrm{cr}}e\bb{u}\times\bb{B}$ term
in the MHD equations, and thus relies on using an appropriate value for
$n_{\mathrm{cr}}/n_i$, where $n_i$ is the number density of ions 
in the plasma. 
A conservative upper limit on this value can be obtained by assuming 
energy equipartition; for ions at GeV energies and CRs at TeV 
energies this gives
\begin{equation}
\frac {n_{\mathrm{cr}}}{n_i} \sim 10^{-3}~
\left( \frac{T_i} {\mathrm{GeV}} \right)
\left( \frac{T_{\mathrm{cr}}} {\mathrm{TeV}} \right)^{-1},
\end{equation}
where $T_i$ and $T_{\mathrm{cr}}$ are typical energies of the 
ions in the plasma and the CRs driving the instability. 
This number is an overestimate 
if anything and decreases with CR energy. On small scales and 
for injection problems, the low energy CRs are important in
driving the instability and charge separation has an effect,
but at late times the highest energy CRs are important. 
We therefore do not expect charge separation to 
have an effect in growing the turbulence to large scales at late 
times, where the CR energies are very high and 
$n_{\mathrm{cr}}/n_i \ll 0.01$. This is the important regime 
for UHECR acceleration. Based on this, 
we argue that the magnetic field strength continues to grow until the scale of 
the turbulence reaches the Larmor radius of the CRs driving 
the instability and the saturation is set by equation~\ref{eq:bsat}. 
In a real astrophysical situation, dynamical timescales
start to affect this conjecture.

\subsection{Amplification timescale}
\label{sec:timescale}
The amplification timescale is important for a number 
of reasons and in different astrophysical systems; in general,
it must be shorter than the various dynamical timescales so that 
CRs in the vicinity of the shock can be efficiently scattered.
In SNRs, one obvious consideration is that the amplification timescale
must be significantly smaller than the age of the SNR; this is known as the 
\cite{LC1983a,LC1983b} limit. Furthermore, in the upstream region it is necessary
that the amplification in a given volume of plasma occurs before the 
shock overtakes it \citep{bell2013}. 

Since the parameters of our simulations are chosen to give a maximum
growth rate roughly equal to $1$, the time $t$ taken for amplification 
by a factor 100 is approximately given by \citep{bell2004,bell2013}
\begin{equation}
\gamma_{\mathrm{max}} t \sim N_{\mathrm{100}},
\end{equation}
where $N_{\mathrm{100}}$ is the normalised time at which amplification by a factor of 
100 has occurred; \cite{bell2013} adopted $N_{\mathrm{100}} \approx 5$
for the parallel case. The fact that the transition to the non-linear 
regime occurs at $\Delta B / B_0 \approx 1$ in 
perpendicular shocks instead results in an increased value 
of $N_{\mathrm{100}} \approx 25$ (which can be inferred from 
Fig.~\ref{fig:b_long}). The location of this transition point 
(see section~\ref{sec:transition}) is therefore
important in estimating the amplification timescale. It is notable 
that the transition occurs at different values of $B$ in each of the
3D perpendicular, 3D parallel and 2D perpendicular cases, emphasising the 
need for a 3D treatment of each configuration. The general arguments presented by 
\cite{bell2004} are unaffected by the increased value of $N_{100}$; 
however, it does suggest that the maximum CR energy attainable 
in perpendicular shocks in SNRs may be lower by a factor $\sim10$ compared
to parallel shocks \cite[see][equation~27]{bell2004}.

\subsubsection{Radio hotspots}

In perpendicular shocks, the NRH instability must 
amplify the magnetic field to $\sim100\mu$G in the time $t_\perp$ 
in which the downstream plasma travels a distance $r_{g0}$, 
where $r_{g0}=T_{\mathrm{max}}/(ecB_0)$ is the Larmor radius of a 
CR with energy $T_{\mathrm{max}}$ spiralling in the {\em unamplified} 
field. This constraint, that $\gamma_{\mathrm{max}} t_\perp \gtrsim 25$,
can be applied to the hotspots of radio galaxies if the shock 
there is perpendicular. Here, we follow \cite{araudo2016a} and 
Araudo et al. (in preparation),
and assume that the magnetic field responsible for
scattering the synchrotron electrons is amplified by the NRH instability.
If we consider the downstream region of a jet hotspot, with initial
ordered field $B_0=B_j$ and number density $n_j$,
we can derive a lower limit on the acceleration efficiency of 
the CRs driving the instability. This acceleration efficiency 
is defined as $\eta = U_{\mathrm{cr}} / U_{\mathrm{kin}}$. 
The current in this region is given by 
\begin{equation}
j \approx n_{\mathrm{cr}} e v_d = e v_d \frac{\eta m_p n_j u_s^2}
{T_{\mathrm{nrh}}},
\end{equation}
where $T_{\mathrm{nrh}}$ is the energy density of the CRs
driving the instability on this scale.
In order to accelerate particles up to an energy $T_{\mathrm{max}}$, 
the mean free path of these particles must be less than their Larmor 
radius \citep{bell2004,kirk2010,sironi2013}. This condition is satisfied
when the turbulence scale size is greater than $(T_{\mathrm{max}}/ecB)^2/r_{g0}$.
Assuming that turbulence grows to the Larmor radius, $r_{g0}$, 
of the driving CRs gives $T_{\mathrm{max}}B_j=T_{\mathrm{nrh}}B$.
If we now adopt some canonical values with a shock 
velocity of $u_s=c/10$, and set the CR drift velocity and plasma drift 
velocities to $u_s$ and $u_s/4$ respectively, we obtain 
\begin{equation}
\eta_{\mathrm{min}} \sim 0.01 \left( \frac{B_j}{\mu G} \right)^2
\left( \frac{B}{100 \mu G} \right)^{-1}
\left( \frac{N_{100}}{25} \right)
\left( \frac{u_s}{c/10} \right)^{-2},
\end{equation}
where we have also introduced an additional 
correction factor to account for the increased value of $N_{100}$,
and adopted the same jet density, 
$n_j=10^{-4}~\mathrm{cm}^{-3}$ as \cite{araudo2016a}. 
\cite{blandford1987} estimate an acceleration efficiency
of a few per cent in SNRs, but the actual value is likely much
higher, at tens of per cent, since the particle acceleration
is a self-regulating process. Regardless, our lower limit is below 
even the more conservative value. 
This suggests that the NRH instability
can amplify magnetic fields in perpendicular shocks 
to the required values on a short enough
timescale, such that $\gamma_{\mathrm{max}} t_\perp \gtrsim 25$.
Future observational efforts will help to constrain these values
further, but the fact that numbers comparable to those observed 
are obtained even for this reduced growth rate suggests that the picture
of particle acceleration is at least consistent and that the values adopted by 
\cite{araudo2016a} are reasonable.

\subsubsection{Further implications}
Our work has more general implications for a number of 
areas where perpendicular shocks are likely important 
in accelerating particles. In SNRs, both perpendicular and 
parallel shocks are thought to accelerate non-thermal 
electrons \citep[e.g.][]{west2017}, and differences
in amplification timescales between the two can be applied to 
models of the synchrotron emission. 
Differences in growth rate turnovers may actually help to 
say something about the magnetic field orientation in 
particle acceleration sites (if the NRH instability is responsible). 
This could prove interesting in radio galaxies, where magnetic field 
orientations are difficult to constrain, yet enter calculations of the scattering efficiency across the shock \citep{reynolds1996,croston2009}.
Furthermore, the fact that turnover to the non-linear growth rate occurs
earlier in perpendicular shocks has implications for models
of CR acceleration, particularly in relativistic shocks
such as those seen in GRBs 
\citep[e.g.][]{panaitescu2002,yost2003,milos2006}
and radio galaxies \citep[e.g.][]{laing2002}, where the 
pre-existing fields are quasi-perpendicular. 
The application of the NRH instability to relativistic shocks
will be discussed in detail by Bell et al. (in preparation).

\section{Conclusions}
\label{sec:summary}
We have studied the NRH instability, in which a $-\jcb$ force from a 
return current acting in opposition to the CR current produces 
purely growing modes of plasma turbulence, 
amplifying the magnetic field. Our results confirm
those of \cite{bell2004} and RS10, in that we find the 
perpendicular NRH instability causes exponential growth of the 
magnetic field strength and energy density in the plasma. 
Our main conclusions are as follows.

\begin{enumerate}
\item Our 3D simulations exhibit a different
plasma structure relative to the 2D simulations, 
as thermal and magnetic pressure terms along
the $z$ axis cause a decreased density contrast.
\item Despite these differences, the instability
continues to grow exponentially in 3D and the non-linear growth 
rate is very similar ($\gamma \approx 0.07j$).
\item At late times, the 3D perpendicular simulation shows 
similarities in structure to the parallel case; in general,
the late-time behaviour is characterised 
by having no preferred
direction to the turbulence, a result only observable 
in 3D simulations.
\item A Rayleigh-Taylor-like analysis can be applied to 
the perpendicular NRH instability and used to describe the linear and early
non-linear regimes.
\item The transition to the non-linear regime is governed by 
thermal pressure and occurs at a different values of $\Delta B/B_0$
in each of the geometries.
This value, denoted $\chi$, 
depends on the critical scale at which pressure 
terms become in important; this is roughly a mode wavelength when 
$\jperpb$ but a larger `transverse coherence length' when $\jparab$.
\item For $\jperpb$, 
The thermal pressure is fairly insensitive to the initial pressure condition
as small scale shocks emerge due to the $\jcbo$ force, rapidly
heating the plasma. The rapid growth of thermal pressure -- 
which is isotropic -- means there is no physical limit that is
well produced by the 2D simulation.
\item Our 2D MHD simulations
reproduce the behaviour found by RS10 in terms of the structure 
and growth rate of the perpendicular NRH instability. 
\item We find that the 3D simulation does not
saturate in even our longest and largest simulations. In 
an astrophysical situation, we argue that the 
maximum magnetic field strength will be set by either the 
consideration that the scale of the turbulence cannot grow beyond
the Larmor radius of the CRs driving the instability, or 
the limiting dynamical timescale. 
\end{enumerate}
The NRH instability thus remains a promising candidate for 
amplifying magnetic fields to the magnitudes required to 
explain the observed non-thermal emission and CR energy spectrum.
Our work also indicates that it can operate regardless of the magnetic field
orientation relative to the cosmic ray current. 
However, further 3D treatments of 
the instability are needed to study its behaviour in further detail.

\section*{Acknowledgements}

We would like to thank John Kirk, Will Potter, Alex Schekochihin
and Andrey Beresnyak for useful discussions. 
This work is supported by the Science and 
Technology Facilities Council 
under grants ST/K00106X/1 and ST/N000919/1.
We thank the referee, Prof. Luke Drury, for his constructive 
comments. We would like to acknowledge the use of the 
University of Oxford Advanced Research Computing (ARC) 
facility in carrying out this work 
(\url{http://dx.doi.org/10.5281/zenodo.22558}).
Simulations were conducted using \textsc{mh3d} version 2.0.0.
Figures are produced using the {\tt matplotlib 2.0} plotting 
library \citep{matplotlib, matplotlib2}.




\bibliographystyle{mnras}
\bibliography{paper} 


\bsp	
\label{lastpage}
\end{document}